\documentclass[a4paper,fleqn,usenatbib]{mnras}

\usepackage{newtxtext,newtxmath}

\usepackage[T1]{fontenc}
\usepackage{ae,aecompl}

\usepackage{graphicx} 
\usepackage{amsmath}  
\usepackage{amssymb}  

\usepackage{amsfonts}
\usepackage{xspace}
\usepackage{gensymb}
\usepackage{subfigure}
\usepackage[usenames, dvipsnames]{color}
\usepackage{pdflscape}

\usepackage{diagbox}
\usepackage{times}
\usepackage[encapsulated]{CJK}
\usepackage[utf8x]{inputenc}

\newcommand{\co}{CO\xspace}
\newcommand{\coonezero}{\mbox{CO(1-0)}\xspace}
\newcommand{\cotwoone}{\mbox{CO(2-1)}\xspace}

\newcommand{\htwo}{H$_{2}$\xspace}

\newcommand{\msolpcsq}{M$_{\odot}\,{\rm pc}^{-2}$\xspace}

\newcommand{\hi}{{\sc H\,i}\xspace}

\newcommand{\hii}{{\sc H\,ii}\xspace}

\title[Local Group Dust Power Spectra]{Spatial Power Spectra of Dust across the Local Group: No Constraint on Disc Scale Height}

\author[Koch et al.]{Eric W. Koch$^{1}$\thanks{E-mail:
koch.eric.w@gmail.com (EWK)}, I-Da Chiang \begin{CJK*}{UTF8}{bkai}(江宜達)\end{CJK*}$^{2}$, Dyas Utomo$^{3}$, J\'er\'emy Chastenet$^{2}$,
\newauthor Adam K. Leroy$^{3}$, Erik W. Rosolowsky$^{1}$, Karin M. Sandstrom$^{2}$\\
$^{1}$University of Alberta, Department of Physics, 4-183 CCIS, Edmonton AB T6G 2E1, Canada\\
$^{2}$Center for Astrophysics and Space Sciences, Department of Physics, University of California, San Diego, 9500 Gilman Drive, La Jolla, CA 92093, USA\\
$^{3}$The Ohio State University, Department of Astronomy, 140 West 18th Avenue, Columbus, OH 43210, USA\\
}

\begin{document}

\date{Draft date: \today}

\pagerange{\pageref{firstpage}--\pageref{lastpage}} \pubyear{2019}

\maketitle

\label{firstpage}

\begin{abstract}
We analyze the 1D spatial power spectra of dust surface density and mid to far-infrared emission at $24\mbox{--}500\,\mu$m in the LMC, SMC, M31, and M33.
By forward-modelling the point-spread-function (PSF) on the power spectrum, we find that nearly all power spectra have a single power-law and point source component.
A broken power-law model is only favoured for the LMC 24~$\mu$m MIPS power spectrum and is due to intense dust heating in 30 Doradus.
We also test for local power spectrum variations by splitting the LMC and SMC maps into $820$~pc boxes.
We find significant variations in the power-law index with no strong evidence for breaks.
The lack of a ubiquitous break suggests that the spatial power spectrum does not constrain the disc scale height.
This contradicts claims of a break where the turbulent motion changes from 3D to 2D.
The power spectrum indices in the LMC, SMC, and M31 are similar ($2.0\mbox{--}2.5$).
M33 has a flatter power spectrum ($1.3$), similar to more distant spiral galaxies with a centrally-concentrated \htwo distribution.
We compare the
power spectra of \hi, CO, and dust in M31 and M33, and
find that \hi power spectra are consistently flatter than CO power spectra.
These results cast doubt on the idea that the spatial power spectrum traces large scale turbulent motion in nearby galaxies.
Instead, we find that the spatial power spectrum is influenced by (1) the PSF on scales below $\sim3$ times the FWHM, (2) bright compact regions (30 Doradus), and (3) the global morphology of the tracer (an exponential CO disc).
\end{abstract}

\begin{keywords}
galaxies: individual (LMC, SMC, M31, M33) --- galaxies: ISM --- ISM: dust --- methods: statistical
\end{keywords}

\section{Introduction}
\label{sec:intro}

Turbulence is an integral part of the dynamics in the interstellar medium (ISM).  Within the inertial range of turbulence, the self-similar structure of the density and velocity fields produce a power-law distribution, which can be measured using statistical techniques like the power spectrum \citep{Elmegreen2004ARA&A..42..211E}.
Together, the density and velocity fields constrain the energy power spectrum $\mathbf{E(k)}$.
This can directly be compared to turbulence models for incompressible \citep{Kolmogorov1941DoSSR..32...16K} and compressible gas \citep{BURGERS1948171,Fleck1996ApJ...458..739F,GaltierBanerjee2011PhRvL.107m4501G,Federrath2013MNRAS.436.1245F}.
ISM observations provide usable constraints on 3D turbulent velocity and density fluctuations from the 2D line-of-sight velocity and column density maps \citep{FederrathRoman-Duval2010A&A...512A..81F}. This connection offers a method for constraining the turbulent energy power spectrum from observational data.

Of particular interest for the star formation process and galaxy evolution is distinguishing what mechanism drives turbulence throughout a galaxy.
Because turbulence decays quickly ($\sim10$~Myr), the ubiquity of observed turbulent properties implies the need for a near-continuous source of turbulent energy injection \citep{MacLowKlessen2004RvMP...76..125M}.
Observational constraints on the turbulent driving scale may provide a clean measurement to distinguish between different sources of energy injection.
This connection can be difficult to make with Milky Way observations as
line-of-sight confusion makes it difficult to distinguish scales at and above the disc scale height ($>100$~pc). 
As a result, high dynamic range extragalactic observations may offer the best way to trace the scale of energy injection.

The spatial power spectrum of a turbulent cascade offers a potential solution to constrain the disc scale height and driving scale in face-on galaxies.  The index of the energy power spectrum changes with both the type of turbulence and the number of spatial dimensions.  For the latter, the index is expected to steepen by $+1$ as the turbulent motions transition from being confined from three- to two-dimensions \citep[e.g.,][]{Lazarian2000ApJ...537..720L}.  Extragalactic observations that resolve scales below the disc scale height are ideal for testing whether this ``break'' in the power spectrum indeed occurs, using the column density or the line-of-sight velocity fields.  From this break scale, the disc scale height can be measured, constraining quantities like the turbulent energy injection on galactic scales \citep[e.g.,][]{Tamburro2009AJ....137.4424T,Koch2018MNRAS,Utomo2019ApJ...871...17U}, and the mid-plane pressure
\citep{Blitz2006ApJ...650..933B} that are used in star formation models based on vertical dynamical equilibrium \citep{Ostriker2010ApJ...721..975O}.

Several studies, primarily using column density or intensity maps, investigate the spatial power spectrum in nearby galaxies.  Some studies find power spectra well-described by both a single power-law \citep[e.g.,][]{Stanimirovic2000MNRAS.315..791S,Dutta2013NewA...19...89D,Zhang2012ApJ...754...29Z}.
Others find a broken power-law \citep[e.g.,][]{Elmegreen2001ApJ...548..749E,Dutta2009MNRAS.397L..60D,Combes2012A&A...539A..67C}, where the break has been interpreted as the disc scale height. 
These and other studies also find a large range in the power law index. This is true even when comparing results use a single traced like the 21-cm \hi line \citep[e.g.,][]{Dutta2013NewA...19...89D}.

This range in extragalactic power spectrum properties makes it difficult to draw general conclusions about the nearby galaxy population.
One reason for the confusion may be that extragalactic power spectrum analyses use heterogeneous data and techniques.
In general, extragalactic studes have also not corrected for steepening on small scales due to the PSF response \citep[excepting][]{Muller2004ApJ...616..845M}, though this effect is commonly account for in Galactic power spectrum analyses \citep{MivilleDeschenes2003A&A...411..109M,Martin2015ApJ...809..153M,Blagrave2017ApJ...834..126B}. This issue was also noted by \citet{Grisdale2017MNRAS.466.1093G} who found that the break points in \hi power spectra in a few nearby galaxies are consistently limited by the PSF scale.

A further issue to consider with extragalactic power spectra is how galactic structure not dominated by turbulence (i.e., spiral arms) affects the power spectrum shape.  These large-scale distributions are known to contribute additional power on large-scales. \citet{Grisdale2017MNRAS.466.1093G} show that changes in the mass distribution steepens the column density power spectrum from galaxy-scale simulations.
\citet{Koch2019MNRAS.485.2324K} show how the clustering of GMC locations in M33's inner disc contributes to an excess in the power spectrum up to scales near the disc scale length \citep[$\sim2$~kpc;][]{Druard2014A&A...567A.118D}.

Accurate measurements of the power spectrum are particularly important now because recent advances in galaxy-scale numerical simulation resolve
similar scales to current observations of Local Group galaxies \citep[e.g.,][]{Grisdale2017MNRAS.466.1093G,Dobbs2018MNRAS.478.3793D,Garrison-KimmelHopkins2019MNRAS.487.1380G}.
Comparing the power spectra between these observations and simulations can provide a powerful diagnostic for how
large-scale galactic structure affects the power spectrum shape \citep[e.g.,][]{Grisdale2017MNRAS.466.1093G}.  For example, several simulations show a power spectrum break that is altered by stellar feedback \citep{Bournaud2010MNRAS.409.1088B,Pilkington2011MNRAS.417.2891P,Combes2012A&A...539A..67C,Grisdale2017MNRAS.466.1093G}, though the prominence of spiral arms also appears to play a role \citep{Renaud2013MNRAS.436.1836R}.

In this paper, we present a uniform analysis of 1D dust emission power spectra in four Local Group galaxies, the Large and Small Magellanic Clouds, M31, and M33.
We use archival {\it Spitzer} and {\it Herschel} data, as well as dust surface density maps from \citet{Utomo2019ApJ...874..141U}.
We compare power spectrum properties across different galactic environments while resolving scale similar to or below the disc scale height \citep[$\sim100$~pc;][]{Kalberla2009ARA&A..47...27K}.
Our analysis models the point-spread function (PSF) on the power spectrum shape and demonstrates that a single power-law combined with unresolved point sources can reproduce most of the observed power spectra.
We present the maps used in \S\ref{sec:observations} and the power spectrum model in \S\ref{sec:power_spectra}.
We discuss the implications of our modelling in \S\ref{sec:discussion}, including comparisons between IR bands and galaxies, and how the dust power spectrum relates to power spectra of \hi, tracing the atomic ISM, and CO, tracing the molecular ISM.
Our uniform power spectrum analysis of multiple phases in multiple galaxies offers a benchmark for simulations of Local Group-like galaxies.

Throughout this paper, we define $P(k)$ as the 1D power spectrum produced from an intensity or surface density maps and the power spectrum index $\beta$ as $\propto k^{-\beta}$ such that $\beta > 0$.

\section{Observations}
\label{sec:observations}

We focus our study on the Magellanic Clouds, M31, and M33. These are the closest targets uniformly observed across the mid- to far-infrared by both {\it Spitzer} \citep{Werner2004ApJS..154....1W}  and {\it Herschel} \citep{Pilbratt2010A&A...518L...1P}. Due to their large angular size and proximity ($<1$~Mpc), these targets maximize the spatial range that can be studied in their power spectra. The {\it Spitzer} and {\it Herschel} maps of the Magellanic Clouds have resolve $\sim10$~pc scale, well below the expected scale height of both the atomic and molecular gas discs.

We use existing {\it Spitzer} MIPS \citep[$24$, $70$, \& $160$~$\mu$m][]{RiekeYoung2004ApJS..154...25R}, and {\it Herschel} PACS \citep[$100$ \& $160$~$\mu$m][]{PoglitschWaelkens2010A&A...518L...2P} and SPIRE \citep[$250$, $350$, \& $500$~$\mu$m][]{GriffinAbergel2010A&A...518L...3G} data products from several projects: LMC {\it Spitzer} MIPS \citep[SAGE;][]{Meixner2006AJ....132.2268M}, {\it Herschel} PACS \& SPIRE \citep[HERITAGE;][]{Meixner2013AJ....146...62M}; SMC {\it Spitzer} MIPS \citep[SAGE-SMC \& S$^3$MC;][]{Gordon2006ApJ...638L..87G,Gordon2011AJ....142..102G,Bolatto2007ApJ...655..212B}, {\it Herschel} PACS \& SPIRE \citep[HERITAGE;][]{Meixner2013AJ....146...62M}; M31 {\it Spitzer} MIPS \citep{Barmby2006ApJ...650L..45B}, {\it Herschel} PACS \& SPIRE \citep{Groves2012MNRAS.426..892G,Draine2014ApJ...780..172D}; M33 {\it Spitzer} MIPS \citep{Hinz2004ApJS..154..259H,Tabatabaei2007A&A...466..509T}, {\it Herschel} PACS \& SPIRE \citep[HerM33es;][]{Kramer2010A&A...518L..67K}.

Altogether, we create 1D power spectra from maps in eight infrared bands in our analysis. We include both the MIPS and PACS $160$~$\mu$m maps to check for consistency between different instrumental effects and noise levels. As a check, we did rerun our analysis on background-subtracted maps and found that the background remove had little effect on the power spectrum properties. This lack of change in the power spectrum is expected since the background tends to be both low intensity and smooth on large-scales.

A key component in our analysis is the effect of the instrumental PSF response on the power spectrum shape. We use the PSF and convolution kernels from \citet{Aniano2011PASP..123.1218A} to model for PSF effects. We also convolve each map to the ``moderate'' Gaussian size provided in Table 6 of \citet{Aniano2011PASP..123.1218A}, again using their publicly available convolution kernels. 


We also analyze the dust surface density maps from \citet{Utomo2019ApJ...874..141U}, which were derived applying a uniform analysis to the \textit{Herschel} data for each of our targets.
A modified blackbody model is fit to the spectral energy distribution (SED) from $100\mbox{-}500$~$\mu$m following the methodology of \citet{Chiang2018ApJ...865..117C} and \citet{Gordon2014ApJ...797...85G}. 
The dust maps are provided at a common set of physical resolutions; here, we use the highest resolution for each galaxy: $13$~pc for the LMC and SMC, and $167$~pc for M31 and M33.

\begin{figure*}
\includegraphics[width=\textwidth]{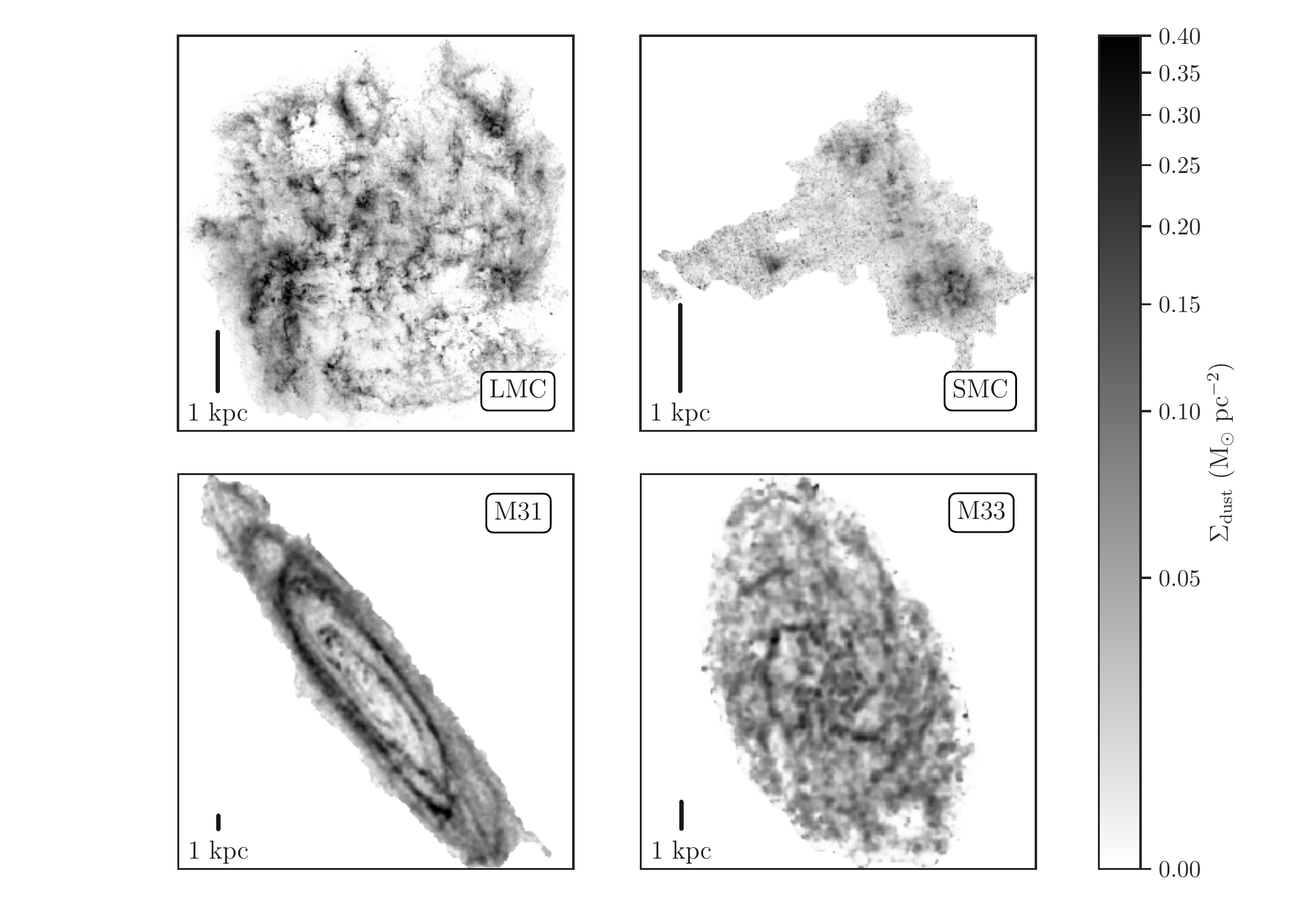}
\caption{\label{fig:coldens_maps} Dust surface density maps of the LMC, SMC, M31, and M33 from \citet{Utomo2019ApJ...874..141U}.  The bars in the lower left corners show the $1$~kpc scale shown for each galaxy. The region shown for each galaxy is used for all images in the power spectrum analysis (\S\ref{sec:power_spectra}).}
\end{figure*}

Figure \ref{fig:coldens_maps} shows the dust surface density maps from \citet{Utomo2019ApJ...874..141U} for each galaxy.
The region displayed in the figure shows the area used for the analysis of all maps in all bands.
Thus, the power spectra can be compared directly.

\pagebreak
Throughout this paper, we adopt distances of 62.1~kpc for the SMC \citep{Graczyk2014ApJ...780...59G}, 50.2~kpc to the LMC \citep{Klein2014arXiv1405.1035K}, 744~kpc to M31 \citep{Vilardell2010A&A...509A..70V}, and 840~kpc to M33 \citep{Freedman2001ApJ...553...47F}. These are the same distances used by \citet{Utomo2019ApJ...874..141U} to create the dust maps at common physical resolutions that we use here.

\section{Power spectrum analysis}
\label{sec:power_spectra}

We characterize and compare the spatial structure in the LMC, SMC, M31, and M33 with the 1D spatial power spectrum from intensity or dust surface density maps, a commonly-used technique for describing ISM structure from $\sim$AU to kpc scales \citep[e.g.,][]{Elmegreen2004ARA&A..42..211E}.
We present the power spectrum calculation in \S\ref{sub:calculating_power_spectra}, the power-law model and fitted results in \S\ref{sub:modelling_power_spectra}, and the model selection criteria in \S\ref{sub:model_selection}. 
Except for the MIPS 24~$\mu$m results for the LMC, all of the power spectra the we measure are well-fit by a single power-law plus point source component.
In \S\ref{sub:lmc_24um}, we demonstrate that 30 Doradus is responsible for a power spectrum break in the LMC MIPS $24$~$\mu$m.
Finally, \S\ref{sub:does_the_power_spectrum_index_change_on_} presents local dust surface density power spectra from $820\times820$~pc$^2$ square regions in the LMC and SMC.
This analyses allows us to explore variations in the power spectrum index.
These local power spectra are also well-fit by a single power-law model.

\subsection{Calculating power spectra}
\label{sub:calculating_power_spectra}

We use {\tt TurbuStat} \citep{Koch2019AJ....158....1K}\footnote{Version 1.0; \url{turbustat.readthedocs.io}} to compute the 1D spatial power spectrum.
{\tt TurbuStat} implements a common version of many turbulence statistics described in the literature, including the spatial power spectrum.
While {\tt TurbuStat} can model the full 2D power spectra of images, we focus this study on modelling 1D power spectra azimuthally-averaged in Fourier space. The 1D power spectrum, $P(k)$, is most commonly used in extra-galactic studies.

When large values are at the edge of the map, the Gibbs phenomenon causes ringing in the Fourier transform, which manifests as a strong cross-shape in the 2D power spectrum. Since the ringing will affect the 1D power spectrum shape, we apply a Tukey function to smoothly taper the edges of the map prior to computing the power spectrum. The maps that require this added step are described in \S\ref{sub:modelling_power_spectra} and Appendix \ref{app:additional_systematics}.

\subsection{Modelling the power spectra}
\label{sub:modelling_power_spectra}

We consider two models to describe the 1D power spectrum shape: (1) a single power-law and (2) a broken power-law. In both cases, we allow an optional point source component.  Both models account for extended emission with the power-law components and the response of bright, individual point sources on small scales with a constant component.


The single power-law model for a 1D power spectrum $P(k)$ is:
\begin{equation}
    \label{eq:model}
    P_{\rm single}(k) = P_{\rm ext}(k) + P_{\rm pt}(k) = A\,k^{-\beta} + B.
\end{equation}
This model has three free parameters to fit: the power-law amplitude $A$, the index $\beta$, and the point-source contribution $B$.

The broken power-law model accounts for a change in the power-law index at some scale. This model has been used in previous extragalactic studies \citep{Block2010ApJ...718L...1B,Combes2012A&A...539A..67C}.  We adopt a broken power-law model following the form implemented in \citet{astropy}:
\begin{equation}
    \label{eq:brokplaw_model}
    P_{\rm broken}(k) = A\,\left( \frac{k}{k_b} \right) ^ {-\beta} \left\{ \frac{1}{2} \left[ 1 + \left( \frac{k}{k_b}\right)^{1 / \Delta} \right] \right\}^{(\beta - \beta_2) \Delta} \, + B.
\end{equation}
This model adds three additional parameters relative to Equation \ref{eq:model}, two of which are left as free parameters when fitting. The parameters $\beta$ and $\beta_2$ describe the power-law index below and above the break $k_b$, respectively. This form of a broken power-law smoothly varies between the power-law components, with the ``smoothness'' set by the $\Delta$ parameter. We fix $\Delta=0.1$ based on visually comparing model solutions.
Given that the fitting is done in frequency pixel units and the bin size of $1$ is used for all power spectra, we expect this to be an appropriate choice for our analysis.
This smooth version of a broken power-law offers a more realistic description of the data, rather than a model with a sharp break at $k_b$

Equations \ref{eq:model} and \ref{eq:brokplaw_model} are physically-motivated, idealized models that do not account for any real observational effects.
In the simplest interpretation, the power-law component results direction from turbulent density fluctuation while the point source component reflect, e.g., young stellar objects and embedded star forming regions
\citep[][]{Seale2009ApJ...699..150S}.

Instrumental systematics affect the observed shape of the power spectrum, causing it to deviate from the idealized models above.  Fortunately, most of these effects can be account for by forward modelling.
In this analysis, we forward model the point spread function (PSF) response for each map. Multiplying by the PSF response ($P_{\rm PSF}(k)$), the models from Equations \ref{eq:model} \& \ref{eq:brokplaw_model} become:
\begin{equation}
    \label{eq:obs_model}
    P_{\rm obs}(k) = P_{\rm PSF}(k)\,\cdot\,P_{\rm model}(k).
\end{equation}
Since the PSF response has a fixed form, it does not introduce additional free parameters in the model.  For a Gaussian response, $P_{\rm PSF}(k) \, \propto \, {\rm exp}\left( - 4\pi^2 \sigma_{\rm beam}^2 k^2  \right)$, where $\sigma_{\rm beam} = {\rm FWHM} / \sqrt{8 {\rm log} 2}$ is the Gaussian rms of the beam.  Similar models that include the PSF response have been used in several studies \citep[e.g.,][]{MD2002A&A...393..749M,Muller2004ApJ...616..845M,Martin2015ApJ...809..153M,Blagrave2017ApJ...834..126B}.

We fit the power spectra of the maps at two resolutions: (1) the native resolution, and (2) convolved to a
Gaussian beam using the ``moderate'' kernels listed in Table 6 of \citet{Aniano2011PASP..123.1218A}. At the native resolution, we account for the non-Gaussian PSF shape by regridding the PSF map from \citet{Aniano2011PASP..123.1218A} to have the same pixel size as the observed map and using its 1D power spectrum as $P_{\rm PSF}(k)$ in Equation \ref{eq:obs_model}. For the convolved maps, we use the analytic form for a Gaussian PSF. 

We found that fits to the the power spectra of the dust surface density maps were improved by including an uncorrelated white noise term $C$:
\begin{equation}
    \label{eq:obs_coldens_model}
    P_{\rm obs}(k) = P_{\rm PSF}(k)\,\cdot\,P_{\rm model}(k) + C.
\end{equation}
This additional $C$ term is due to fitting the dust SED to individual pixels.
The inherent uncertainty of the SED fit adds some noise to the dust surface density map.
Since the fits are performed for each pixel, this additional noise is not affected by the PSF and is reasonably approximated as white (uncorrelated) noise.


Since the amplitudes $A$, $B$ and $C$ in the model vary over several orders of magnitude, we fit the $\log_{10}$ of these parameters to make it easier to sample large variations.  Due to the potentially wide range in parameters from map to map, we adopt uninformative uniform priors on the parameters:
\begin{align}
    \label{eq:priors}
    \log_{10} \, A \, &\sim \, \mathcal{U}\left(-20, 20\right)\\
    \log_{10} \, B \, &\sim \, \mathcal{U}\left(-20, 20\right)\\
    \log_{10} \, C \, &\sim \, \mathcal{U}\left(-20, 20\right) && \text{for Eq. \ref{eq:obs_coldens_model}}\\
    \beta \, &\sim \, \mathcal{U}\left(0, 10\right)\\
    \beta_2 \, &\sim \, \beta + \mathcal{N}\left(0, 10\right) && \text{for Eq. \ref{eq:brokplaw_model}}\\
    k_b \, &\sim \, \mathcal{U}\left(k_{\rm min}, k_{\rm max}\right) && \text{for Eq. \ref{eq:brokplaw_model}}
\end{align}
The chosen parameter ranges are significantly wider than the expected values and none of the fitted parameters converged to the edge of a parameter range. We also note that adopting wide Gaussian priors on the parameters did not affect the fits.

When fitting Equation \ref{eq:brokplaw_model}, we treat the second power-law component, on scales below $k_b$, as a perturbation on the large-scale index $\beta$.  This allows for $\beta\sim\beta_2$, thereby converging to Equation \ref{eq:model} when a break is not strongly preferred in the fit.  The break point $k_b$ is sampled uniformly over the whole range of spatial frequencies $k$. The importance of $k_b$ diminishes at large $k_b$ where forward-modelling the PSF response dominates the power spectrum shape.  In these cases, $\beta_2$ and $k_b$ could be well-constrained but the fit will be indistinguishable from Equation \ref{eq:model}.

We fit the 1D power spectra and assume that the standard deviation of the azimuthal average is a reasonable uncertainty.
Since most of these maps have a high signal to noise ratio, particularly on large scales, the variations in radial bins will be larger than the inherent uncertainty.
Thus, we treat the 1D power spectrum values ($P_{\rm 1D}$) as independent samples drawn from a normal distribution with a width inferred from the standard deviation in azimuthal bins.
We draw these samples in log$_{10}$ space to avoid sampling negative values for the power spectrum:
\begin{equation}
    \label{eq:data_normal}
    \mathrm{log}_{10} P_{\rm 1D}^{\star} \, \sim \, \mathcal{N} \left(\mathrm{log}_{10} P_{\rm 1D}, \sigma_{\mathrm{log}_{10} P_{\rm 1D}}  \right).
\end{equation}
The model is fit to the sampled values $P_{\rm 1D}^{\star}$, ensuring the data uncertainty is reflected in the parameter posterior distributions.

When fitting maps from the photometric bands to Equation \ref{eq:obs_model}, we only consider scales above $3\times$ the Gaussian standard deviation of the beam to avoid regions where pixelization or convolution residuals dominate the power spectrum shape (see Appendix \ref{app:additional_figures}).
We also limit the fit to scales less than $k < k_{\rm max}/3$, where $k_{\rm max}$ is the inverse of half the map shape. This removes large deviations in the largest bin that arise from the need to account for the total intensity in the image. These large scale bins are estimated from just a few samples in the 2D power spectrum and thus have a large uncertainty (see Figure \ref{fig:coldens_fits}).  The dust surface density maps from \citet{Utomo2019ApJ...874..141U} have pixel sizes $2.5$ times smaller than the beam which avoids small scales dominated by convolution residuals. Therefore, we include the smallest scales of the column density power spectra in the fit to the surface density.

We use the {\sc pymc3} package \citep{pymc3} to fit the models, using Sequential Monte Carlo to sample the parameter space \citep{DelMoral-2006-smc}, as we found it rapidly converged for this problem. Sequential Monte Carlo runs a set of parallel Markov chains through a series of stages. At each stage, the sampling progresses from the prior to posterior distribution by tempering, controlled by a tempering parameter\footnote{Typically $\beta$ is used for the tempering parameter, but we adopt $\beta^\star$ to avoid confusion with the power spectrum index $\beta$.} $\beta^\star$. At each stage, $\beta^\star$ is increased according to the samples in the previous step, starting at $0$ for the prior distribution and ending at $1$ for the posterior distribution. For our fits, we found a good balance between computational cost and convergence using 100 chains that sample over 6000 iterations for each step.  For comparison, we fit several power spectra using the Levenberg-Marquardt algorithm, which provides similar parameter values but severely underestimates parameter uncertainties and covariance. Using Markov Chain Monte Carlo (MCMC) and accounting for the data uncertainty (Equation \ref{eq:data_normal}) provides realistic parameter uncertainties.

\subsection{Model Selection}
\label{sub:model_selection}

We fit each of the power spectra to the single (Equation \ref{eq:model}) and broken power-law (Equation \ref{eq:brokplaw_model}) models while forward-modelling the PSF response. To compare the models, we compute the Widely-applicable Information Criterion \citep[WAIC;][]{Watanabe2010_waic}, as implemented in {\sc pymc3}, to determine the preferred model. WAIC estimates the out-of-sample prediction accuracy from a Bayesian model based on the log-likelihood from the MCMC parameter samples, with a correction for the number of variables to account for overfitting.  We note that the model comparisons calculated using leave-one-out (LOO) cross-validation are consistent with those from WAIC for our fits \citep{Vehtari2017_waic_loo}.

We choose the preferred models by comparing the WAIC and its uncertainty between the two models.  The preferred model should minimize the WAIC. However, we find that the WAIC is similar for many of the fits. In this case, we choose the simpler single power-law model (Equation \ref{eq:model}) given the lack of clear evidence for a broken power-law.  In many of these cases, the break point approaches the PSF FWHM and the broken power-law has a diminishing influence on the fit quality.

\subsection{Fit Results}
\label{sub:results}

For all of the fits but one (\S\ref{sub:lmc_24um}), we find that the power spectra are well-fit by a single power-law and point source model (Equation \ref{eq:model}) with no significant requirement for a broken power-law.
In our measurements, the PSF response can account for any observed steepening of the power spectra on smaller scales.

\begin{figure*}
\includegraphics[width=0.9\textwidth]{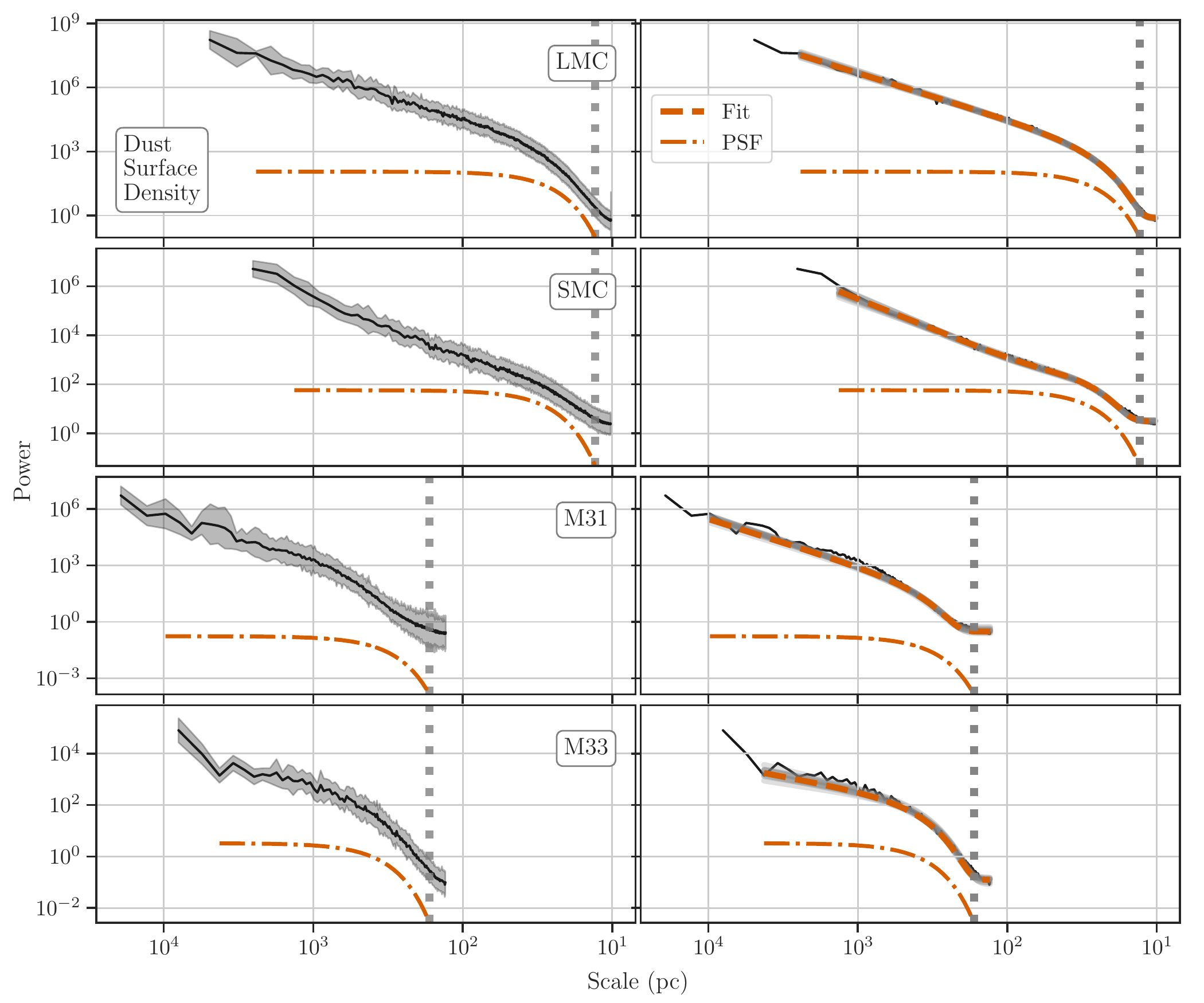}
\caption{\label{fig:coldens_fits} 1D power spectra of the dust surface density (left column) and the fitted models (right column; Table \ref{tab:coldens_fits}) for the LMC, SMC, M31, and M33 shown on a common physical scale. The orange dashed-dotted line shows the power spectrum of the PSF response, scaled to an order of magnitude below the amplitude from the fit. The vertical thick gray dotted line is the FWHM of the PSF response. The standard deviation on the power spectrum is shown in the shaded gray region in the left panel.  The dashed orange line in the right column panels is the best fit model, and the underlying gray lines are the model fits from 10 random draws of the MCMC.  The power spectra are well-fit by a single power-law and point-source component, while the PSF response is solely responsible for the break point on small scales.}
\end{figure*}

Figure \ref{fig:coldens_fits} shows the power spectra and fits for the dust surface density maps from \citet{Utomo2019ApJ...874..141U}. The PSF response, shown separately for each galaxy, has a noticeable effect on the shape of the power spectrum on scales $\sim3\mbox{--}4$ times the FWHM. By incorporating the PSF response into our model, the fits shown in Figure \ref{fig:coldens_fits} account for the apparent break point on those scales.  Table \ref{tab:coldens_fits} provides the fitted parameters using Equation \ref{eq:obs_coldens_model}.  These results show that the power spectra are all well-fit by a single power-law, plus a point source term for maps with $\sim10$~pc resolution (i.e., the LMC and SMC) and do not require a physical break point in the model.

\begin{table*}
    \caption{\label{tab:coldens_fits} Fit parameters for the 1D dust surface density power spectra to Equation \ref{eq:obs_coldens_model}. Uncertainties are the 1-$\sigma$ interval estimated from the MCMC samples. Missing entries in ${\rm log}_{10} B$ are unconstrained in the fit and not used.}
    \centering
    \begin{tabular}{ccccccc}
Galaxy & Resolution ($\arcsec$) & Phys. Resolution (pc) & log$_{10}$ $A$ & $\beta$ & log$_{10}$ B & log$_{10}$ C \\  \hline 
LMC & 53.4 & 13 & $1.67\pm0.08$ & $2.18\pm0.05$ & $2.82\pm0.32$ & $0.79\pm0.08$ \\ 
SMC & 43.2 & 13 & $-0.16\pm0.25$ & $2.47\pm0.15$ & $2.52\pm0.04$ & $2.93\pm0.18$ \\ 
M31 & 46.3 & 167 & $0.02\pm0.13$ & $2.46\pm0.14$ & -- & $0.33\pm0.06$ \\ 
M33 & 41.0 & 167 & $1.32\pm0.15$ & $1.11\pm0.14$ & -- & $0.13\pm0.03$ \\
\end{tabular}

\end{table*}

\begin{figure*}
\includegraphics[width=0.9\textwidth]{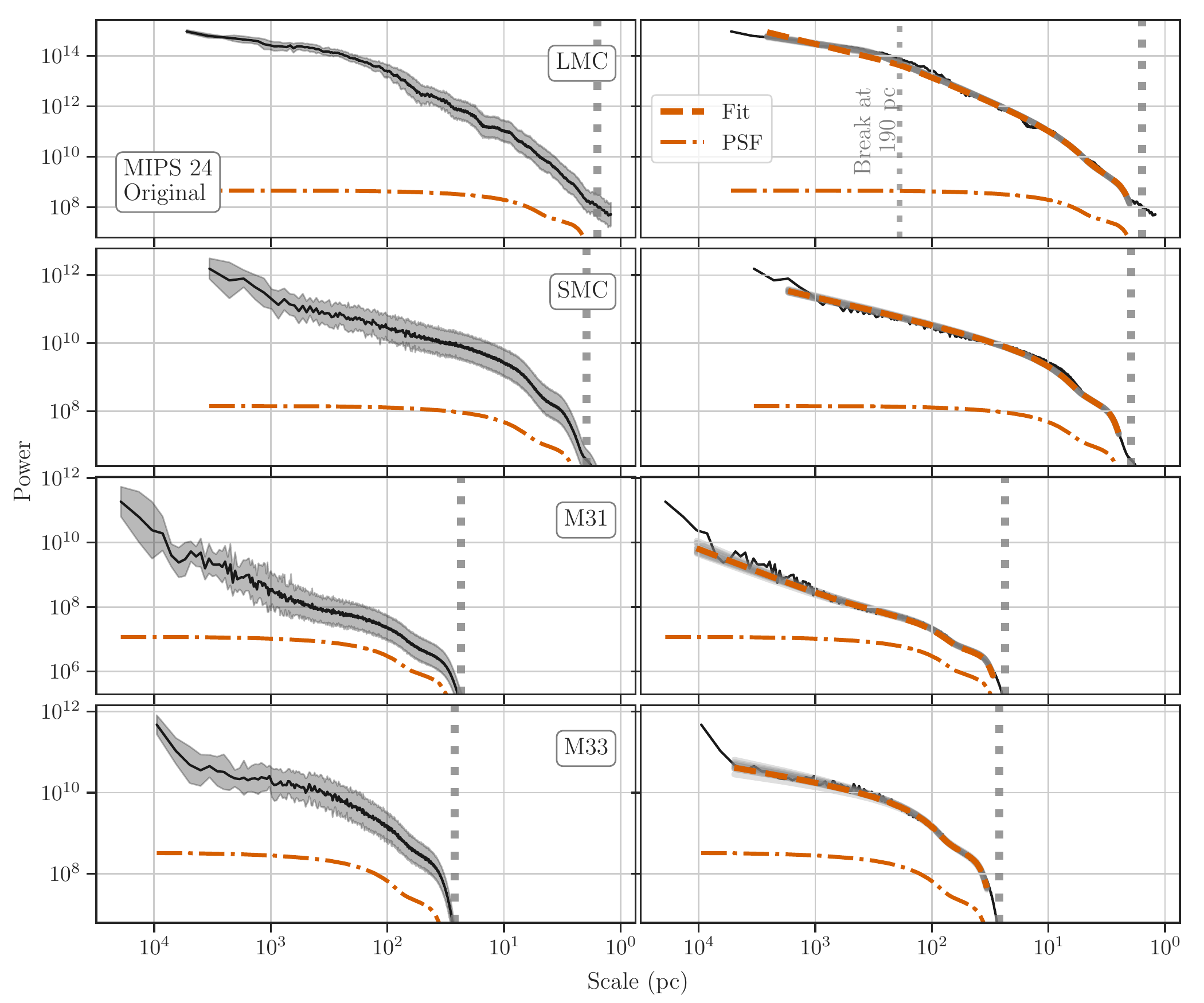}
\caption{\label{fig:mips24_fits} 1D power spectra of MIPS $24$~$\mu$m maps (left column) and the fitted models (right column; Table \ref{tab:band_fits}), similar to Figure \ref{fig:coldens_fits}.  A similar figure for each band is included as supplemental information. The MIPS $24$~$\mu$m PSF is highly non-Gaussian, as is clear from the extra ``step'' in the response on small scales. Similar to the dust surface density, the power spectra are well fit without a break point. The only exception across all of the bands is the LMC MIPS 24~$\mu$m power spectrum in the top panel. A broken power-law model (Eq. \ref{eq:brokplaw_model}) is marginally preferred with a break point at $190$~pc, shown in the labeled gray vertical line.  We show in \S\ref{sub:lmc_24um} that 30 Doradus causes this deviation.}
\end{figure*}

Figure \ref{fig:mips24_fits} shows the power spectra and fits for the MIPS $24$~$\mu$m maps at their original resolution. The MIPS $24$~$\mu$m PSF has noticeable non-Gaussian features that result in a ``step'' in the PSF response\footnote{For M33, the step in the PSF is on $\sim80$~pc scales.}.
Each of the galaxy maps show this step feature in their power spectra, indicating these breaks are solely from the PSF.  All of the power spectrum fits are best described by a single power-law model, except for the LMC MIPS 24~$\mu$m shown in the top panel of Figure \ref{fig:mips24_fits}, which we explore in further detail in \S\ref{sub:lmc_24um}.   Repeating this analysis with the MIPS $24$~$\mu$m convolved to a Gaussian PSF, we find consistent power-law indices, demonstrating that the fits are not strongly dependent on the PSF model.  Table \ref{tab:band_fits} provides the complete fit parameters for all bands and resolutions, with the WAIC used for model selection.

\begin{table*}
    \caption{\label{tab:band_fits} Fit parameters for the individual band power spectra to Equations \ref{eq:model} \& \ref{eq:brokplaw_model} with forward-modelling the PSF response (Equation \ref{eq:obs_model}. Uncertainties are the 1-$\sigma$ interval estimated from the MCMC samples. Missing entries in ${\rm log}_{10} B$ are unconstrained in the fit and not used, while missing entries in other parameters ($\beta_2$ and $k_b$) are cases where the broken power-law model is not preferred. The model preference is indicated by the {\bf bold-faced} WAIC value and chosen based on a $1\sigma$ difference for the broken power-law, or $\leq1\sigma$ for the single power-law.  The latter is chosen based on the lack of evidence for a more complex model.  Fits labelled with $^{\bigstar}$ do not include the PSF shape in the model (see Appendix \ref{app:additional_systematics}), and those labelled with $^{\blacklozenge}$ use an apodizing kernel to minimize ringing in the FFT from large values at the image edges.}
    \centering
    \begin{tabular}{ccccccccccc}
 & & FWHM & FWHM & & & & & $x_b=1 / k_b$ & BP & SP \\
Galaxy & Band & ($\arcsec$) & (pc) & log$_{10}$ $A$ & $\beta$ & log$_{10}$ $B$ & $\beta_2$ & (pc) & WAIC ($\times10^3$) & WAIC ($\times10^3$) \\  \hline
LMC & MIPS 24 &  6.5 &  2 & $13.65\pm0.09$ & $1.13\pm0.03$ & -- & $1.79\pm0.07$ & $190\pm25$ & $\mathbf{241.0\pm0.4}$ & $241.5\pm0.4$ \\
    &      &  11.0 &  3 & $14.28\pm0.07$ & $0.50\pm0.09$ & -- & $1.77\pm0.18$ & $439\pm46$ & $\mathbf{152.9\pm0.3}$ & $153.2\pm0.3$ \\
    & MIPS 70 &  18.7 &  5 & $10.14\pm0.04$ & $1.70\pm0.02$ & $10.96\pm1.40$ & -- & -- & $102.2\pm0.2$ & $\mathbf{102.2\pm0.2}$ \\  
     &      &  30.0 &  7 & $10.22\pm0.03$ & $1.67\pm0.02$ & -- & -- & -- & $67.1\pm0.2$ & $\mathbf{67.1\pm0.2}$ \\
    & PACS 100 &  7.1 &  2 & $11.17\pm0.04$ & $1.73\pm0.02$ & $13.13\pm0.02$ & -- & -- & $289.6\pm0.3$ & $\mathbf{289.7\pm0.3}$ \\  
     &      &  9.0 &  2 & $11.24\pm0.04$ & $1.70\pm0.02$ & $13.02\pm0.02$ & -- & -- & $233.7\pm0.3$ & $\mathbf{233.7\pm0.3}$ \\
    & MIPS 160 &  38.8 &  9 & $7.84\pm0.07$ & $2.13\pm0.04$ & $9.05\pm0.71$ & -- & -- & $40.7\pm0.2$ & $\mathbf{40.8\pm0.02}$ \\
     &      &  64.0 &  16 & $8.07\pm0.05$ & $2.00\pm0.04$ & -- & -- & -- & $26.3\pm0.1$ & $\mathbf{26.3\pm0.1}$ \\
    & PACS 160 &  11.2 &  3 & $10.07\pm0.04$ & $1.88\pm0.02$ & $11.61\pm0.04$ & -- & -- & $166.3\pm0.3$ & $\mathbf{166.3\pm0.3}$ \\
     &      &  14.0 &  3 & $10.14\pm0.05$ & $1.84\pm0.03$ & $11.45\pm0.08$ & -- & -- & $137.5\pm0.3$ & $\mathbf{137.5\pm0.3}$ \\
    & SPIRE 250 &  18.2 &  4 & $8.59\pm0.06$ & $1.89\pm0.04$ & -- & -- & -- & $89.7\pm0.2$ & $\mathbf{89.7\pm0.2}$ \\
     &      &  21.0 &  5 & $8.64\pm0.04$ & $1.87\pm0.03$ & -- & -- & -- & $79.2\pm0.2$ & $\mathbf{79.2\pm0.02}$ \\
    & SPIRE 350 &  25 &  6 & $7.31\pm0.03$ & $1.95\pm0.02$ & -- & -- & -- & $56.5\pm0.2$ & $\mathbf{56.5\pm0.2}$ \\
     &      &  28.0 &  7 & $7.33\pm0.03$ & $1.95\pm0.03$ & -- & -- & -- & $51.3\pm0.2$ & $\mathbf{51.3\pm0.2}$ \\
    & SPIRE 500 &  36.4 &  9 & $6.16\pm0.03$ & $1.99\pm0.03$ & -- & -- & -- & $33.9\pm0.2$ & $\mathbf{33.9\pm0.2}$ \\
     &      &  41.0 &  10 & $6.18\pm0.04$ & $1.97\pm0.03$ & -- & -- & -- & $30.7\pm0.2$ & $\mathbf{30.7\pm0.2}$ \\
SMC & MIPS 24 &  6.5 &  2 & $8.91\pm0.02$ & $0.78\pm0.01$ & -- & -- & -- & $116.2\pm0.2$ & $\mathbf{116.2\pm0.2}$ \\
     &      &  11.0 &  3 & $9.09\pm0.03$ & $0.67\pm0.02$ & -- & -- & -- & $72.7\pm0.2$ & $\mathbf{72.7\pm0.2}$ \\
    & MIPS 70$^{\bigstar}$ &  18.7 &  6 & $7.17\pm0.33$ & $1.98\pm0.14$ & $10.56\pm0.38$ & -- & -- & $144.5\pm0.4$ & $\mathbf{144.5\pm0.4}$ \\
     &      &  30.0 &  9 & $7.51\pm0.37$ & $1.83\pm0.17$ & $10.67\pm0.23$ & -- & -- & $145.9\pm0.4$ & $\mathbf{145.9\pm0.4}$ \\
    & PACS 100 &  7.1 &  2 & $8.47\pm0.16$ & $1.93\pm0.07$ & $11.78\pm0.01$ & -- & -- & $131.1\pm0.2$ & $\mathbf{131.1\pm0.2}$ \\
     &      &  9.0 &  3 & $8.68\pm0.16$ & $1.85\pm0.07$ & $11.74\pm0.01$ & -- & -- & $105.5\pm0.2$ & $\mathbf{105.5\pm0.2}$ \\
    & MIPS 160$^{\blacklozenge}$ &  38.8 &  12 & $5.76\pm0.14$ & $2.33\pm0.10$ & $7.51\pm0.08$ & -- & -- & $16.4\pm0.1$ & $\mathbf{16.4\pm0.1}$ \\  
     &      &  64.0 &  19 & $6.03\pm0.23$ & $2.15\pm0.15$ & -- & -- & -- & $10.7\pm0.1$ & $\mathbf{10.7\pm0.1}$ \\
    & PACS 160$^{\bigstar,\blacklozenge}$ &  11.2 &  3 & $7.40\pm0.06$ & $2.16\pm0.04$ & -- & -- & -- & $37.4\pm0.1$ & $\mathbf{37.4\pm0.1}$ \\
     &      &  14.0 &  4 & $7.74\pm0.05$ & $2.04\pm0.03$ & -- & -- & -- & $38.0\pm0.1$ & $\mathbf{38.0\pm0.1}$ \\
    & SPIRE 250 &  18.2 &  5 & $6.24\pm0.11$ & $2.17\pm0.07$ & $8.19\pm0.04$ & -- & -- & $37.2\pm0.2$ & $\mathbf{37.2\pm0.2}$ \\  
     &      &  21.0 &  6 & $6.27\pm0.13$ & $2.15\pm0.08$ & $8.17\pm0.06$ & -- & -- & $32.9\pm0.2$ & $\mathbf{32.9\pm0.2}$ \\
    & SPIRE 350 &  25 &  8 & $5.05\pm0.12$ & $2.28\pm0.08$ & $6.88\pm0.05$ & -- & -- & $23.1\pm0.1$ & $\mathbf{23.1\pm0.1}$ \\
     &      &  28.0 &  8 & $5.11\pm0.11$ & $2.24\pm0.07$ & $6.87\pm0.06$ & -- & -- & $20.9\pm0.1$ & $\mathbf{20.9\pm0.1}$ \\
    & SPIRE 500 &  36.4 &  11 & $3.88\pm0.15$ & $2.39\pm0.09$ & $5.93\pm0.06$ & -- & -- & $13.6\pm0.1$ & $\mathbf{13.6\pm0.1}$ \\  
     &      &  41.0 &  12 & $3.98\pm0.15$ & $2.32\pm0.10$ & $5.90\pm0.07$ & -- & -- & $12.3\pm0.1$ & $\mathbf{12.3\pm0.1}$ \\
M31 & MIPS 24 &  6.5 &  23 & $5.55\pm0.34$ & $1.39\pm0.18$ & $7.82\pm0.68$ & -- & -- & $50.9\pm0.1$ & $\mathbf{50.9\pm0.1}$ \\
     &      &  11.0 &  40 & $5.08\pm0.35$ & $1.59\pm0.16$ & $7.90\pm0.50$ & -- & -- & $31.7\pm0.1$ & $\mathbf{31.7\pm0.1}$ \\
    & MIPS 70 &  18.7 &  67 & $4.74\pm0.42$ & $2.11\pm0.20$ & $7.52\pm0.08$ & -- & -- & $16.5\pm0.1$ & $\mathbf{16.5\pm0.1}$ \\
     &      &  30.0 &  108 & $4.59\pm0.44$ & $2.16\pm0.22$ & $7.53\pm0.13$ & -- & -- & $10.9\pm0.1$ & $\mathbf{11.0\pm0.1}$ \\
    & PACS 100$^{\bigstar}$ &  7.1 &  26 & $8.34\pm0.13$ & $1.91\pm0.07$ & -- & -- & -- & $19.9\pm0.1$ & $\mathbf{19.9\pm0.1}$ \\
     &      &  9.0 &  32 & $8.77\pm0.14$ & $1.77\pm0.07$ & -- & -- & -- & $20.2\pm0.1$ & $\mathbf{20.2\pm0.1}$ \\
    & MIPS 160$^{\blacklozenge}$ &  38.8 &  140 & $5.79\pm0.23$ & $1.87\pm0.19$ & -- & -- & -- & $8.05\pm0.08$ & $\mathbf{8.05\pm0.08}$ \\
     &      &  64.0 &  231 & $5.76\pm0.16$ & $1.89\pm0.16$ & -- & -- & -- & $5.21\pm0.08$ & $\mathbf{5.21\pm0.08}$ \\
    & PACS 160$^{\blacklozenge}$ &  11.2 &  40 & $7.95\pm0.18$ & $2.20\pm0.08$ & $11.57\pm0.04$ & -- & -- & $43.3\pm0.1$ & $\mathbf{43.3\pm0.1}$ \\
     &      &  14.0 &  50 & $8.18\pm0.21$ & $2.11\pm0.09$ & $11.44\pm0.07$ & -- & -- & $35.5\pm0.1$ & $\mathbf{35.5\pm0.1}$ \\
    & SPIRE 250$^{\blacklozenge}$ &  18.2 &  66 & $6.10\pm0.07$ & $2.16\pm0.07$ & -- & -- & -- & $18.4\pm0.1$ & $\mathbf{18.4\pm0.1}$ \\
     &      &  21.0 &  76 & $6.10\pm0.08$ & $2.16\pm0.07$ & -- & -- & -- & $16.4\pm0.1$ & $\mathbf{16.4\pm0.1}$ \\
    & SPIRE 350$^{\blacklozenge}$ &  25 &  90 & $5.01\pm0.07$ & $2.21\pm0.08$ & -- & -- & -- & $11.5\pm0.1$ & $\mathbf{11.5\pm0.1}$ \\  
     &      &  28.0 &  101 & $5.03\pm0.08$ & $2.21\pm0.08$ & -- & -- & -- & $10.5\pm0.1$ & $\mathbf{10.5\pm0.1}$ \\
    & SPIRE 500$^{\blacklozenge}$ &  36.4 &  131 & $4.04\pm0.11$ & $2.18\pm0.11$ & -- & -- & -- & $6.8\pm0.1$ & $\mathbf{6.8\pm0.1}$ \\
     &      &  41.0 &  148 & $4.07\pm0.12$ & $2.17\pm0.12$ & -- & -- & -- & $6.2\pm0.1$ & $\mathbf{6.2\pm0.1}$ \\
M33 & MIPS 24 &  6.5 &  26 & $9.25\pm0.09$ & $0.46\pm0.05$ & -- & -- & -- & $25.7\pm0.1$ & $\mathbf{25.7\pm0.1}$ \\
     &      &  11.0 &  45 & $9.26\pm0.09$ & $0.46\pm0.05$ & -- & -- & -- & $15.9\pm0.1$ & $\mathbf{15.9\pm0.1}$ \\
    & MIPS 70 &  18.7 &  76 & $8.21\pm0.08$ & $0.99\pm0.06$ & -- & -- & -- & $8.38\pm0.06$ & $\mathbf{8.38\pm0.06}$ \\
     &      &  30.0 &  122 & $8.30\pm0.11$ & $0.94\pm0.08$ & -- & -- & -- & $5.51\pm0.05$ & $\mathbf{5.51\pm0.05}$ \\
    & PACS 100 &  7.1 &  29 & $9.24\pm0.25$ & $1.30\pm0.13$ & -- & -- & -- & $26.7\pm0.1$ & $\mathbf{26.7\pm0.1}$ \\
     &      &  9.0 &  37 & $9.47\pm0.10$ & $1.20\pm0.06$ & -- & -- & -- & $21.7\pm0.1$ & $\mathbf{21.7\pm0.1}$ \\
    & MIPS 160 &  38.8 &  158 & $6.18\pm0.11$ & $1.50\pm0.11$ & -- & -- & -- & $3.19\pm0.05$ & $\mathbf{3.19\pm0.05}$ \\
     &      &  64.0 &  261 & $6.43\pm0.28$ & $1.30\pm0.21$ & -- & -- & -- & $2.06\pm0.03$ & $\mathbf{2.06\pm0.03}$ \\
    & PACS 160 &  11.2 &  46 & $8.61\pm0.09$ & $1.31\pm0.07$ & -- & -- & -- & $15.6\pm0.1$ & $\mathbf{15.6\pm0.1}$ \\
     &      &  14.0 &  57 & $8.52\pm0.09$ & $1.37\pm0.06$ & -- & -- & -- & $12.9\pm0.1$ & $\mathbf{12.9\pm0.1}$ \\
    & SPIRE 250 &  18.2 &  74 & $6.68\pm0.07$ & $1.51\pm0.06$ & -- & -- & -- & $7.83\pm0.07$ & $\mathbf{7.83\pm0.07}$ \\
     &      &  21.0 &  86 & $6.73\pm0.08$ & $1.47\pm0.07$ & -- & -- & -- & $6.94\pm0.07$ & $\mathbf{6.94\pm0.07}$ \\
    & SPIRE 350 &  25 &  102 & $5.54\pm0.07$ & $1.42\pm0.07$ & -- & -- & -- & $4.75\pm0.06$ & $\mathbf{4.75\pm0.06}$ \\
     &      &  28.0 &  114 & $5.60\pm0.08$ & $1.38\pm0.07$ & -- & -- & -- & $4.32\pm0.07$ & $\mathbf{4.32\pm0.07}$ \\
    & SPIRE 500 &  36.4 &  148 & $4.66\pm0.09$ & $1.17\pm0.08$ & -- & -- & -- & $2.72\pm0.05$ & $\mathbf{2.72\pm0.05}$ \\
     &      &  41.0 &  167 & $4.72\pm0.10$ & $1.15\pm0.08$ & -- & -- & -- & $2.46\pm0.05$ & $\mathbf{2.46\pm0.05}$ \\
\end{tabular}

\end{table*}

In a few cases, we found that the power spectra did not follow the PSF response on small scales. In each case, other systematic effects not included in the model dominate the power spectrum shape. These cases are indicated by a $\blacklozenge$ in Table \ref{tab:band_fits} and a longer explanation is provided in Appendix \ref{app:additional_systematics}.  In most cases, we found that cross-hatching of telescope scans near bright emission enhances the power on scales near to or smaller than the FWHM of the PSF. It is difficult to account for this effect in our model, so we instead fit a model without the PSF response (Equation \ref{eq:model}) and limit the scales fit to several times the PSF FWHM.

\subsection{A break due to 30 Doradus in the LMC MIPS $24$~$\mu$m power spectrum}
\label{sub:lmc_24um}

\begin{figure}
\includegraphics[width=0.5\textwidth]{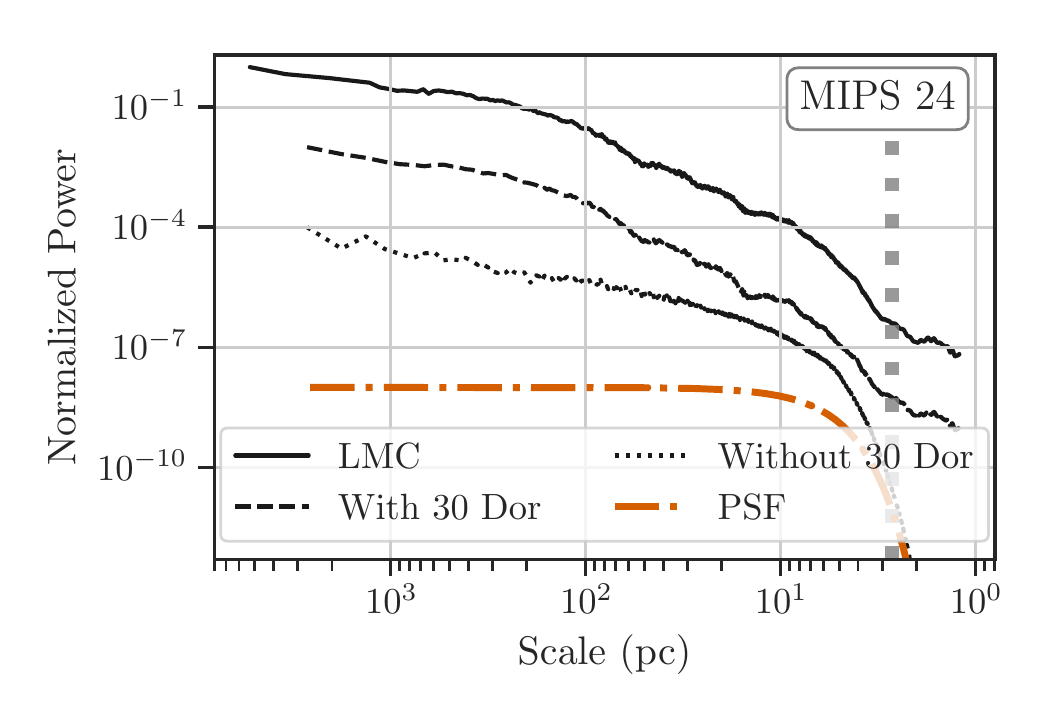}
\caption{\label{fig:lmc_mips24_30dor} power spectra of the MIPS 24~$\mu$m image for the LMC convolved to an $11\arcsec$ ($3$~pc) Gaussian beam, and equal area regions ($4\times3.6$~kpc$^2$ area) that include and exclude 30 Doradus. The power spectra are normalized by their maximum and offset by $10^{2}$ to show the relative shapes.  For the entire LMC and the region including 30 Doradus, there is a deviation and break point from a single power-law at $\sim80$~pc. In the region without 30 Doradus, however, the power spectrum is well-described by a single power-law. In this case, the flux from 30 Doradus relative to the entire LMC is sufficient to produce a break in the power spectrum.}
\end{figure}

We find that all power spectra in our sample are well-fit by a single power-law plus point source model with the exception of the LMC MIPS $24$~$\mu$m map.  For the LMC at $24$~$\mu$m, Figure \ref{fig:lmc_mips24_30dor} shows a distinct bump in the power spectrum on scales of $>80$~pc, making the broken power-law model (Equation \ref{eq:brokplaw_model}) preferred based on the WAIC of each fit.  This feature is also noted by \citet{Block2010ApJ...718L...1B}.  However, the break scale from broken power-law model is not well-constrained.
We derive somewhat different values when we fit the data at their original resolution and when we fit the map after smoothing to a Gaussian kernel (Table \ref{tab:band_fits}).
In this section, we identify 30 Doradus (30 Dor) as the source for this break. 

A significant fraction ($\sim30\%$) of the LMC's emission at $24$~$\mu$m is solely from this giant \hii region. We investigate the effect that a prominent single source has on the power spectrum by calculating the power spectrum in $3$~kpc boxes with and without 30 Dor, where the edges of both boxes have the same apodizing kernel applied to suppress Gibbs ringing.  Figure \ref{fig:lmc_mips24_30dor} shows the distinct difference in these power spectra. The box containing 30 Dor has a power spectrum that closely matches the LMC's power spectrum and contains a similar bump at the same scales. In contrast, the box without 30 Dor follows a power-law to the scale of the box.

Figure \ref{fig:lmc_mips24_30dor} shows the power spectrum from two $3$~kpc boxes, one with and one without 30 Dor.  The region with 30 Dor clearly shows a similar deviation from a single power-law matching the power spectrum of the whole galaxy.  The region without 30 Dor does not show this deviation is an noticably shallower, similar to the MIPS 24~$\mu$m power spectra of the other galaxies (Table \ref{tab:band_fits}).  We fit a single power-law to the power spectrum without 30 Dor and find an index of $1.29\pm0.02$, steeper than the large-scale index of $1.13\pm0.03$ from the power spectrum of the entire image.

We initially fit a broken power-law (Eq. \ref{eq:brokplaw_model}) to the power spectrum with 30 Dor, however, the break point from the fit converges to the beam size.  Restricting the fit to larger scales did not lead to an improved fit. We therefore only fit  a single power-law model to the power spectrum from this region.  The single power-law fit gives an index of $1.76\pm0.02$, which is consistent with the index below the break point from the power spectrum of the entire galaxy.  Since this fit agrees with the entire image power spectrum, the broken power-law fit likely did not converge due to the lack of data points at the largest scales within the $3$~kpc region.  This is similar to the change in the break point between the original and convolved LMC MIPS 24~$\mu$m power spectra.  The break point is very sensitive to the data.



This example demonstrates how the power spectrum shape can be significantly altered by a small number of regions with large intensity relative to the whole image.  \citet{Willett2005AJ....129.2186W} find a similar result in the power spectrum of NGC 2366 in H$\alpha$, where a giant \hii region causes an additional power spectrum ``bump.''  Images with power spectrum breaks should be tested for whether the break is due to a limited number of bright discrete features.

\subsection{Power spectrum variations within local $820$~pc regions} 
\label{sub:does_the_power_spectrum_index_change_on_}

We test whether the power spectrum varies across the LMC and SMC by computing the dust column density power spectra in local ($\sim820$~pc) regions.
We choose 820~pc to balance between measuring local variations and retaining sufficient information to constrain the power spectrum.
The high physical resolution ($\sim10$~pc) allows for a large spatial range to be studied in the local power spectra.
Due to the lower physical resolution in M31 and M33, we cannot access such a large spatial range in those galaxies. 

Previous power spectrum studies of the Magellanic Clouds find spatial variations in the power-law index \citep{Muller2004ApJ...616..845M}, potentially tracing variations in the turbulence.
Furthermore, \citet{Padoan2001ApJ...555L..33P} identify a spatially varying scale height across the LMC studying the \hi emission from $180$~pc regions using the Spectral Correlation Function.

We test for a varying break point by splitting the LMC and SMC dust surface density maps into $\sim820$~pc overlapping squares and fit the power spectra in each region with Equation \ref{eq:obs_coldens_model}. These square regions tend to have bright emission at their edges, so we apply a Tukey apodizing kernel to remove ringing in the FFT, as explained in \S\ref{sub:calculating_power_spectra}.  We focus only on the Magellanic Clouds for this analysis since the resolution of $13$~pc is an order of magnitude below the $\sim$few $100$~pc disc scale height of atomic gas in dwarf galaxies \citep{Walter1999AJ....118..273W}.


There are large signal-to-noise (S/N) variations among the regions. The vast majority of regions have sufficient signal to measure the power law component, though a few regions in the SMC are clearly dominated by noise.  When fitting the local power spectra, we found that the S/N variation leads to strong correlations between the $B$ and $C$ parameters from Equation \ref{eq:obs_coldens_model}. Since this analysis is primarily concerned with variation in the power law properties (index and break point), we limit the fits in this section to scales of $4\times\,{\rm FWHM}\approx52$~pc and only fit the power law (i.e., we fix $B=10^{-20}$, the lower limit on the prior). This spatial limit on the fit still captures the ``break'' due to the PSF response and so is adequate for this analysis.

We further attempted to model the local power spectra with a broken power-law model (Eq. \ref{eq:brokplaw_model}) to test for the presence of strong breaks. However, we do not find any cases where the broken power-law model is strongly preferred.  This is in part due to the smaller spatial range, where there is a lack of data points on $>400$~pc to constrain a break.  We focus the remainder of this analysis on variations in the power spectrum index.

\begin{figure*}
\includegraphics[width=\textwidth]{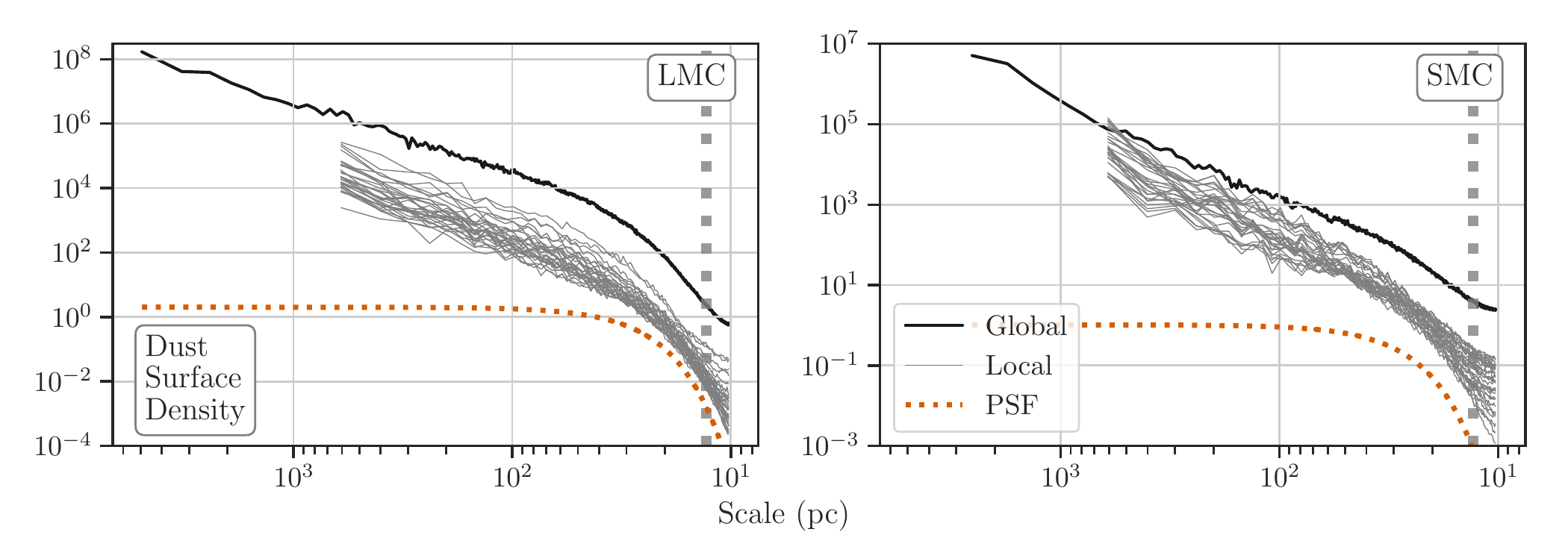}
\caption{\label{fig:magcloud_local_pspecs} Dust surface density power spectra of the entire LMC and SMC maps (black lines) and thirty randomly-selected power spectra (gray lines) from $\sim820$~pc local regions. The dashed line is the PSF scaled to compare with the power spectra shape. Local power spectra steepen at small scales according to the PSF response and do not show a distinct break point on small scales. There is substantial spatial variation in the local power spectrum index, which is shown in Figure \ref{fig:magcloud_index_spatvar}.}
\end{figure*}

Figure \ref{fig:magcloud_local_pspecs} shows LMC and SMC power spectra from the whole map and from randomly-selected $820$~pc boxes. In all cases, we find that the PSF accounts for the power spectrum shape on small scales and that a break point is not required in the model. The lack of a break point on local scales rule out a varying disc scale height as an explanation for finding no break point in the whole galaxy power spectra.

\begin{figure*}
\includegraphics[width=\textwidth]{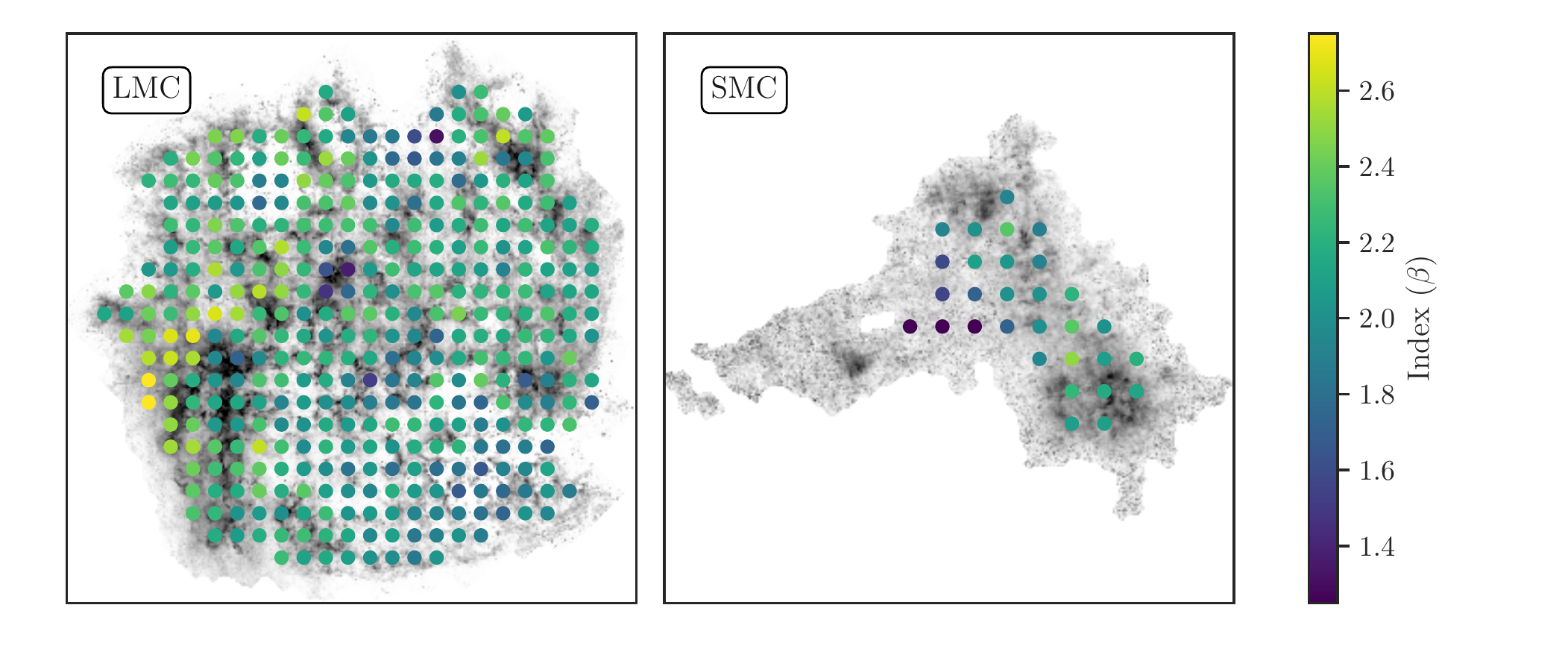}
\caption{\label{fig:magcloud_index_spatvar} The LMC and SMC dust surface density maps overlaid with the power spectrum index measured in $820$~pc regions. There is a $400$~pc overlap between regions so the points shown in the figure are correlated with their neighbours.
We restrict the analysis to high S/N regions, which removes regions at the edge of the LMC and a significant portion of the SMC. In both galaxies, there is significant variation in the power spectrum index (see uncertainty map in Figure \ref{fig:magcloud_index_spatvar_errs}) from the index over the whole galaxy ($2.18\pm0.05$ and $2.47\pm0.15$ for the LMC and SMC, respectively).}
\end{figure*}

The local power spectra in Figure \ref{fig:magcloud_local_pspecs} show variation on scales unaffected by the PSF response, implying that the fitted amplitude and power law index vary with position in both galaxies. While the amplitude is set by the total emission in the box, variations in the index imply changes in the emission morphology. Figure \ref{fig:magcloud_index_spatvar} shows the power spectrum index ($\beta$) overlaid on the dust surface density maps. The power spectrum indices vary from $1.31\pm0.23$ to $2.88\pm0.22$ and $0.82\pm0.38$ to $2.50\pm0.22$ in the LMC and SMC, respectively, after removing all regions near the edges of the column density maps where the noise increases. The range in local power spectrum indices is drastically larger than the uncertainty on the global power spectrum index for both galaxies ($2.18\pm0.05$ and $2.47\pm0.15$; Table \ref{tab:coldens_fits}).  This then implies that the local power spectrum variations are real and not due to noise fluctuations.

In the higher surface density regions to the north and west, the SMC power spectrum varies in index by $\pm1$.
The steepest power spectra have indices of $2.5$ and are offset from the highest surface density regions, suggesting steeper power spectra are sensitive to the gradients in surface density.  In these regions, there is an excess of emission on larger scales and a deficit of small scales---due to the offset from the peak in the emission---leading to a steep power spectrum.  A similar result is found by \citet{Burkhart2010ApJ...708.1204B}, who find extremes in the skewness and kurtosis of the local \hi surface density distribution near large gradients.  The LMC index map shows a similar trend, with steep power spectra offset from 30 Dor and south along the Molecular Ridge along the eastern edge of the LMC.

Spatial variations in the \hi power spectrum or structure function, the real-space analog, have been noted in the Magellanic Bridge \citep{Muller2004ApJ...616..845M} and the SMC \citep{NestingenPalm2017ApJ...845...53N}. In the latter work, \citet{NestingenPalm2017ApJ...845...53N} split the SMC into regions based on the star formation rate.  They find no change in the index with star formation rate, which is somewhat different than the variations correlated with IR brighntess that we observe.
In addition to using a different tracer (they use \hi, we use dust), the regions they use are significantly larger than the $820$~pc boxes than from this analysis. In particular, the bright north and west regions are included in the same high SFR region, which is the area we find moderate variations in the power spectrum index.

More recently, \citet{SzotkowskiYoder2019arXiv191104370S} use the ``rolling power spectrum'' to explore changes in the \hi power spectrum with spatial position in the SMC and LMC. They find evidence of power spectrum breaks only in the LMC, where the power spectra flatten above the break.  Since we do not find this behaviour over the same regions using the dust column or IR bands, this suggests that the \hi may be better coupled to stellar feedback than the total gas column traced by the dust.  While \citet{SzotkowskiYoder2019arXiv191104370S} do not forward model the PSF response, the power spectra are cut-off at the beam scale.  This implies that breaks on scales much larger than the beam ($3\times {\rm FWHM} \sim 90$~pc) are robust against the PSF shape.

\section{Discussion}
\label{sec:discussion}

We show that the IR and dust surface density power spectra for the LMC, SMC, M31 and M33 are well-modeled by a single power-law with point-source term, when the PSF response is accounted for.  A broken power-law model is only preferred for the LMC MIPS 24~$\mu$m image and results from 30~Doradus (\S\ref{fig:lmc_mips24_30dor}).

Here we discuss trends in the power spectrum properties across bands and galaxies and compare with previous studies, some of which have found evidence for breaks in the power spectrum.
We also compare the dust, CO and \hi power spectra in M31 and M33.
We find discrepancies in the fitted index of these three tracers.
This strongly suggests that a comprehensive spatial power spectrum analysis requires a multi-tracer approach.


\subsection{Comparisons with literature power spectra} 
\label{sub:literature_powerspectra}

This paper uses a large suite of archival observations, many of which have been previously analyzed using the spatial power spectrum.
Here, we present an overview of spatial power spectra in the four galaxies analyzed here, including tracers not explored in this work.
Where appropriate, we compare our results to these previous works, highlighting discrepancies in fit values that occur due to different fitting procedures.
Accounting for differences in methodology, our power spectrum fits agree with previous analyses using similar data sets.
We then compare how the power spectra from the IR and dust surface density compares to literature values at other bands.
We note here that our definition of the power-law index ($\propto k^{-\beta}$) is defined so measured indices should have $\beta>0$.
Where appropriate, we alter the sign of literature values to follow this convention.

\subsubsection{LMC}
\label{subsub:lmc_pspec_lit}
\citet{Block2010ApJ...718L...1B} use the LMC MIPS maps \citep{Meixner2006AJ....132.2268M} fit to a two-component power law model, where the component on large scales should have a similar index to our fits.  On large scales, they find indices of $0.78\pm0.19$, $1.83\pm0.36$, and $2.15\pm0.48$ at 24, 70, and 160~$\mu$m, respectively. The latter two agree with our fitted indices, while the 24~$\mu$m is flatter due to the influence of 30 Dor (\S\ref{fig:lmc_mips24_30dor}).

The power spectrum of \hi in the LMC is presented in \citet{Elmegreen2001ApJ...548..749E} and \citet{Elmegreen2003aApJ...590..271E}, where they find that the large-scale index is around $5/3$ on larger-scales ($>100$~pc), as would be expected for Kolmogorov turbulence. 

\subsubsection{SMC}
\label{subsub:smc_pspec_lit}
In the SMC, \citet{Stanimirovic2000MNRAS.315..791S} present power spectra from 60, and 100~$\mu$m {\it IRAS} bands. When fit to a single power law model, the 60 and 100~$\mu$m power spectra have indices of $2.4\pm0.2$ and $3.2\pm0.3$, respectively.  Small scales that appear to be affected by the PSF response are included in the fit, leading to steeper power spectra than if only the large scales were fit; the 100~$\mu$m with its larger PSF is more affected by the decrease in power on small scales.  Consistent with this difference in the power spectrum models, we find much shallower power spectra of $1.98\pm0.14$ and $1.93\pm0.07$ in the MIPS 70 and 100~$\mu$m bands, respectively.  Accounting for uncertainty, the {\it IRAS} 60~$\mu$m index from \citet{Stanimirovic2000MNRAS.315..791S} is consistent with our fit to the MIPS 70~$\mu$m.

In \hi, \citet{Stanimirovic1999MNRAS.302..417S} find a power spectrum index of $3.04\pm0.02$ \citep[see also][]{Stanimirovic2000MNRAS.315..791S,NestingenPalm2017ApJ...845...53N}, steeper than the power spectra we find for the dust surface density ($2.47\pm0.15$).


\subsubsection{M33}
\label{subsub:m33_pspec_lit}
\citet{Combes2012A&A...539A..67C} present power spectra from a large number of tracers for M33, including the MIPS, PACS, and SPIRE bands. Like the LMC analysis by \citet{Block2010ApJ...718L...1B}, they fit a two-component power law model to the power spectra; we compare the large-scale indices with ours. A second difference between our analyses is the fits from \citet{Combes2012A&A...539A..67C} include the two smallest spatial frequency bins, which we do not include in our fits as the values in these bins are significantly larger than what would be inferred from the power law model.  The indices we find tend to be shallower than those found reported by \citet{Combes2012A&A...539A..67C}, consistent with excluding the smallest frequency bins. The discrepancies are the largest for the MIPS bands, where our fitted indices are $\sim0.5$ smaller. The discrepancy is smaller and roughly within the index uncertainty for the PACS and SPIRE maps.

\citet{Combes2012A&A...539A..67C} include power spectra of the molecular and atomic neutral ISM traced through \hi and \cotwoone, respectively \citep{Gratier2010A&A...522A...3G,Druard2014A&A...567A.118D}.
On large-scales (unaffected by the PSF shape), they find indices of $2.4$ for the \hi and $1.5$ for \cotwoone.
The latter does not flatten on large-scales.
\citet{Koch2019MNRAS.485.2324K} shows that an excess of power on large scales is due to the clustering of GMCs in the inner few~kpc.
We also explore these tracers in M33 in \S\ref{sub:comparisons_with_other_ism_tracers} and compare the recovered indices there.

\citet{Combes2012A&A...539A..67C} also include additional tracers that we do not explore, including H$\alpha$, and {\it GALEX} NUV and FUV.  They find indices of $1.2$ for the NUV and FUV bands, similar to what we find in the IR bands.  The H$\alpha$ index they find is $0.77$, similar to the MIPS 24~$\mu$m power spectrum and consistent with an independent study by \citet{Elmegreen2003bApJ...593..333E}.  These similarities are expected since these are tracers of star formation that inherits some of the galactic ISM structure.


Finally, \citet{Elmegreen2003bApJ...593..333E} present power spectra of M33 in the B, V, and R bands. These optical bands are dominated by the stellar component and are not expected to match the ISM-dominated maps that we explore here.
By calculating a power spectrum from 1D azimuthal strips, they find indices of $0.66\pm0.66$.
The large uncertainty is due to contamination from foreground stars.
 

\subsubsection{M31}
\label{subsub:m31_pspec_lit}

There is little previous work on spatial power spectra in M31, likely due to its high inclination.
We discuss the similarity of M31's power spectra to those of the Magellanic Clouds in \S\ref{sub:galaxy_variations}.

\subsection{Variations in the power spectrum index across bands}
\label{sub:index_variations}

\begin{figure}
\includegraphics[width=0.5\textwidth]{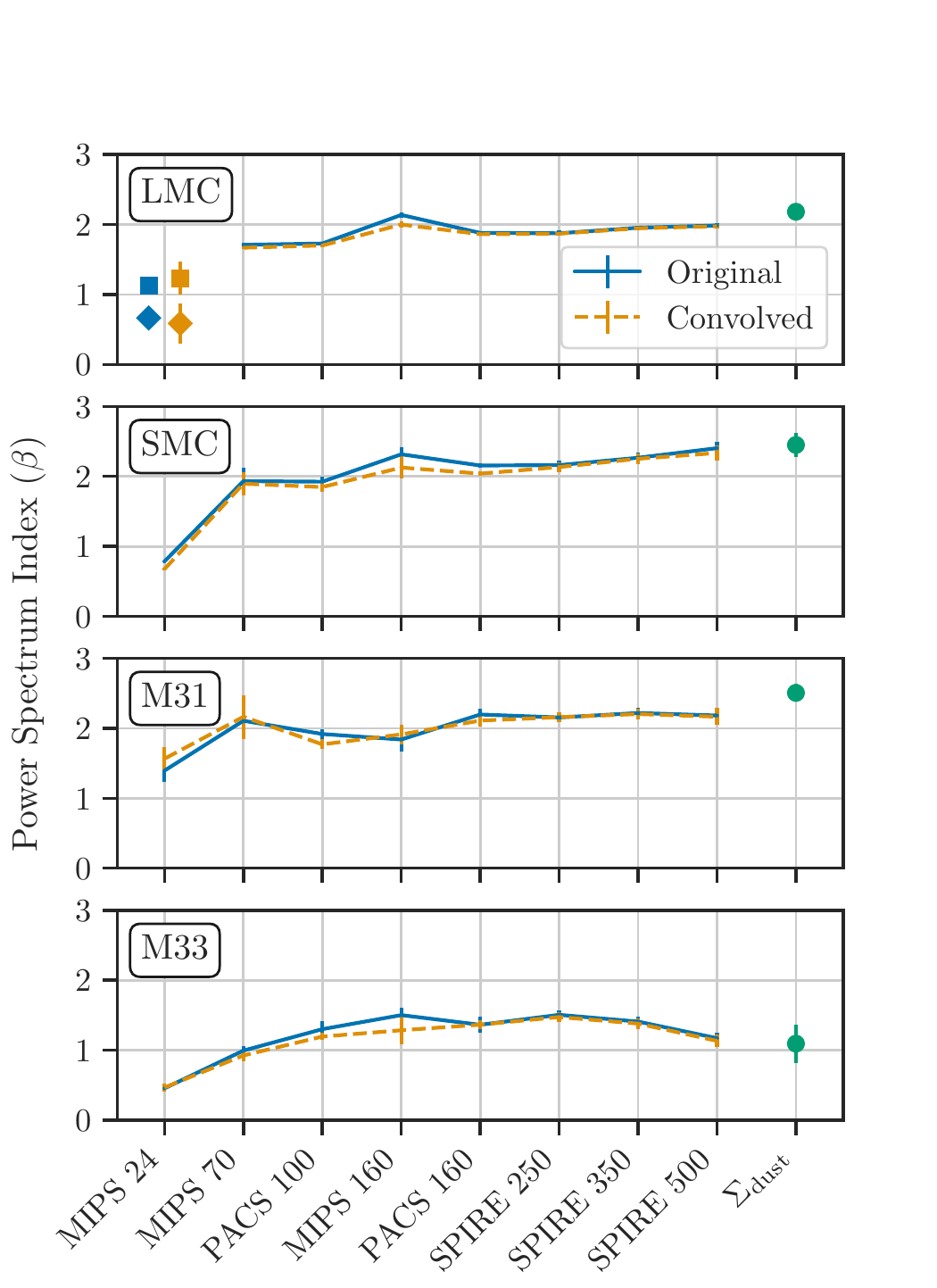}
\caption{\label{fig:pspec_comparison} 1D power spectrum index with $1\sigma$ uncertainty error-bars for each band and the dust surface density map across all four galaxies (see Tables \ref{tab:coldens_fits} \& \ref{tab:band_fits}).  We show the indices for the original images and the images convolved to a Gaussian, using the kernels described in \citet{Aniano2011PASP..123.1218A}, to highlight the agreement between the two results, which is expected if the powers-spectra are well-fit without a break point.  The LMC MIPS $24\,\mu$m has both indices from the broken power-law fit shown, where squares indicate the index on small scales (above the break) and diamonds indicates the index on large-scales (below the break).  The MIPS $24\,\mu$m indices are consistently flatter than the other bands, while all other have slopes of $\sim2$, in broad agreement with literature values. M33 has a consistently flatter power spectrum relative to the other galaxies.}
\end{figure}

Large-scale variations in the shape of the dust spectral energy density, e.g., due to temperature variations, could alter the shape of the power spectrum measured at IR different bands.  Figure \ref{fig:pspec_comparison} summarizes our fitted power spectrum indices from Tables \ref{tab:coldens_fits} \& \ref{tab:band_fits} for each galaxy.  The MIPS 24~$\mu$m index is consistently shallower than those at longer wavelengths by $0.5\mbox{--}1.2$.  As we have shown in \S\ref{sub:lmc_24um}, bright emission regions at $24$~$\mu$m, like 30 Doradus in the LMC, can alter the power spectrum shape over a large range of scales. Bright concentrated sources could result in the shallow power spectra we find in all four galaxies.  The 24~$\mu$m power spectra have a similar index to studies using H$\alpha$ in nearby galaxies, consistent with bright emission in both tracers arising from compact star-forming regions \citep{Elmegreen2003bApJ...593..333E,Combes2012A&A...539A..67C}.

At longer wavelengths, there is less variation in the index. In the LMC and SMC, there is a mild trend of steeper power spectra at longer wavelength.  The power spectrum index in M31 is relatively constant across the longer wavelength bands, while M33's power spectrum in the three SPIRE bands becomes shallower.  With these differences, and the limitations of a small sample size, we find no consistent trend in the power spectrum index from $70\mbox{--}500$~$\mu$m. 



\subsection{Variation in the power spectrum index between galaxies}
\label{sub:galaxy_variations}

The LMC, SMC, and M31 differ significantly in their large-scale morphology (Figure \ref{fig:coldens_maps}), yet they have a similar power spectrum index. Figure \ref{fig:pspec_comparison} shows that the LMC, SMC, and M31 have similar power spectra ranging from $2\mbox{--}2.5$, excluding the MIPS 24~$\mu$m band.
In all cases, the power spectra are shallower than the $8/3$ predicted for 2D Kolmogorov turbulence \cite[e.g.,][]{Elmegreen2004ARA&A..42..211E}.

The small range in power spectrum indices of the LMC, SMC, and M31 could suggest the dust emission shares a similar morphology when resolving $<167$~pc scales, though this similarity is not apparent from the maps in Figure \ref{fig:coldens_maps}.  For the LMC and SMC, where $<13$~pc scales are resolved, bright features in the dust surface density maps correspond to GMCs, and, particularly for the LMC, there are large voids from supershells \citep{Kim1999AJ....118.2797K}. The $167$~pc resolution of the M31 dust surface density map is not sufficient to resolve individual GMCs, and the bright regions in the map primarily highlight M31's ring structure at $R_{\rm gal}\sim10$~kpc. This discrepancy in the spatial morphology at different scales demonstrates that different spatial morphologies can produce similar power spectra.

M33 has a consistently flatter power spectrum compared to the other three galaxies, excluding the MIPS 24~$\mu$m band, with indices ranging from $1\mbox{--}1.5$.  This difference in the power spectrum index results from M33's flocculent spiral morphology with a central enhancement of molecular gas, which differs from the distributed molecular gas morphology in the LMC and SMC, and the predominant molecular rings in M31.  M33's molecular gas, and thus the highest dust surface density, is centrally concentrated into the inner few kpc; \citet{Druard2014A&A...567A.118D} show that the azimuthally-averaged molecular gas surface density, from \cotwoone, is well-fit by an exponential disc with a scale length of $2.1$~kpc. This is in contrast with the \hi distribution, which has a roughly constant average surface density of $\sim8$~\msolpcsq in the inner $8$~kpc \citep{Druard2014A&A...567A.118D,Koch2018MNRAS}.
Approximating the disc as a uniform exponential disc, the power spectrum should have a break near the disc scale length\footnote{The inclination would broaden the power spectrum break.} and a flat power spectrum on larger scales.  While this is a plausible explanation for the \cotwoone power spectrum from \citet{Combes2012A&A...539A..67C} \citep[see][]{Koch2019MNRAS.485.2324K}, this simple model does not explain the lack of a power spectrum break near the disc scale length nor the shallow power spectrum measured for the dust surface density and IR bands.  This implies that the more diffuse and predominantly atomic gas plays an important role in setting the large scale power spectrum shape.  We compare the \hi properties to the dust in M33 in \S\ref{sub:comparisons_with_other_ism_tracers}.

To test this hypothesis of shallow power spectra from a centralized \htwo distribution, we require other nearby face-on galaxies where similar physical scales to the M33 observations can be resolved, of which there are few.
We choose to compare with the face-on spiral galaxy IC 342, which has a distance of $3.4$~kpc \citep{TullyCourtois2013AJ....146...86T}.
Using the Herschel bands \citep{Kennicutt2011PASP..123.1347K}, the PACS 160~$\mu$m resolution resolves $11.2\arcsec\,\approx\,180$~pc scales, similar to the physical resolution of the SPIRE 500~$\mu$m and dust surface density maps of M31 and M33 used here\footnote{There are discrepancies in the expected PSF shape for the IC 342 PACS 100~$\mu$m map and so we exclude it for this comparison.}.


We perform the same analysis on IC 342 that is described in \S\ref{sec:power_spectra}.
We find that IC 342 has a flat power spectrum similar to M33, with typical indices around $1$ (see Table \ref{tab:ic342_fits}), deviating from the index from the LMC, SMC, and M31.
The similarity between M33 and IC 342 suggests that galaxies with a centrally concentrated \htwo-distribution tend to have flatter power spectra, consistent with our expectation above for an exponential (molecular) disc plus a constant (mostly atomic) component.
\citet{Grisdale2017MNRAS.466.1093G} find a similar result in their analysis of power spectra from galaxy-scale simulations and \hi data from THINGS \citep{Walter2008AJ....136.2563W}. They demonstrate this dependence on the gas mass distribution by including an extended uniform gas component with different surface densities and find that this added component steepens the power spectrum on large scales.
The connection with 2D turbulence on large scales is then tenuous for these galaxies.

Though the methodology and resolution of the data differ, in general the power spectrum indices that we find agree with previous work on other galaxies.
Indices from power spectra measured in various optical bands range from $0.6$ to $1.8$ \citep{Elmegreen2003bApJ...593..333E,Willett2005AJ....129.2186W,Elmegreen2006ApJ...644..879E}, while those from $3.6\mbox{--}8.0$~$\mu$m range from $0.8\mbox{--}2.8$ \citep{Block2009ApJ...694..115B}.  The most studied tracer, and with the widest range in indices, is the 21-cm \hi line. Previous studies find indices that range from $0.3$ \citep{Dutta2013NewA...19...89D} to $4.3$ \citep{Zhang2012ApJ...754...29Z}, though most indices range from $1.5\mbox{--}3.0$ \citep{Begum2006MNRAS.372L..33B,Dutta2008MNRAS.384L..34D,Dutta2009MNRAS.397L..60D,Dutta2009bMNRAS.398..887D,Zhang2012ApJ...754...29Z,Dutta2013NewA...19...89D,Dutta2013MNRAS.436L..49D}.  Low-inclination spiral galaxies tend to have flatter power spectra in previous studies \citep{Dutta2013MNRAS.436L..49D}, broadly consistent with our findings for M33 and IC 342.

Spatial power spectra within the Milky Way tend to be steeper than in extragalactic systems, though they around found in a similar range.
The power spectra we find here are consistently shallower than power spectra from Milky Way studies. Galactic \hi power spectra typically have indices from $2.5\mbox{--}4$ \citep{Deshpande2000ApJ...543..227D,Dickey2001ApJ...561..264D,MivilleDeschenes2003A&A...411..109M,Pingel2013ApJ...779...36P,Martin2015ApJ...809..153M,Blagrave2017ApJ...834..126B,Pingel2018ApJ...856..136P}, with extreme values of $2.2$ \citep{Green1993MNRAS.262..327G} to $4.9$ \citep{Kalberla2017A&A...607A..15K}. Values from dust include $2.7$ from extinction over the Perseus molecular cloud \citep{Pingel2018ApJ...856..136P}, $2.9$ from diffuse galactic light in optical bands \citep{MivilleDeschenes2016A&A...593A...4M}, and $2.7$ from Herschel SPIRE maps of the Polaris flare \citep{MivilleDeschenes2010A&A...518L.104M}.  The latter example is $\sim0.5$ steeper than the indices we find for SPIRE maps of the LMC, SMC, and M31 (Table \ref{tab:band_fits}).  We note that Milky Way studies using the spatial power spectrum do not find strong evidence for power spectrum breaks.


    

Even if multiple spatial distributions yield the same power spectrum index, our results still a key benchmark for simulations that aim to reproduce Local Group-like galaxies.
Several recent works aim to simulate galaxies with properties closely matching the LMC, SMC, M31, M33, or the Milky Way \citep{Combes2012A&A...539A..67C,Wetzel2016ApJ...827L..23W,Grisdale2017MNRAS.466.1093G,Dobbs2018MNRAS.478.3793D,Garrison-KimmelHopkins2019MNRAS.487.1380G}  with many producing ``synthetic'' observations to compare with properties found in the actual observations \citep[e.g.,][]{Dobbs2019MNRAS.485.4997D}, a key step for directly comparing simulations and observations \citep{Haworth2018NewAR..82....1H}.
For any simulation the produces dust maps or synthetic IR observations, matching our measured power spectrum represents an important check.

\subsection{Comparisons with \hi and CO power spectra} 
\label{sub:comparisons_with_other_ism_tracers}

The dust surface density closely traces the total neutral gas surface density, related only through the dust-to-gas ratio.  In contrast, 21-cm \hi or CO emission traces only a particular phase of the neutral ISM. This makes the dust surface density a potentially useful tool to compare how the power spectrum changes in the atomic and molecular ISM phases.

We found in the previous section (\S\ref{sub:galaxy_variations}) that the large-scale galactic distribution affects the power spectrum shape. In this section, we compare the dust power spectrum with those from HI and CO, which cleanly separate the atomic and molecular components of the neutral ISM, to measure how the power spectrum changes in the different neutral ISM components.

Due to the different conditions in each neutral ISM phase, the turbulent properties in the \hi and \htwo may differ. \citet{RomeoBurkert2010MNRAS.407.1223R} demonstrate how the transsonic or subsonic conditions in warm \hi alter the stability conditions relative to the supersonic turbulence from \htwo in molecular clouds. In particular, if warm \hi traces a transsonic or subsonic density field, density fluctuations from the mean will be small. This means that the 1D power spectrum from the \hi surface density should be flatter than the \htwo (traced by CO).  These differences in the turbulent properties of the atomic and molecular ISM have important consequences for setting the local stability of the galactic disc \citep{HoffmannRomeo2012MNRAS.425.1511H,RomeoAgertz2014MNRAS.442.1230R}.

Furthermore, dust may be a passive tracer in ISM turbulence, meaning that it may not actively contribute to the turbulence and may have different properties from the gas \citep[e.g.,][]{Goldman2000ApJ...541..701G}. Dust may further be subject to additional drag instabilities \citep{Hopkins2014ApJ...797...59H,HopkinsSquire2018MNRAS.479.4681H}.
The comparisons between the dust, \hi and CO we show here may results from any of these sources. We focus our analysis on looking for consistent differences between the power spectra of these different tracers.



Previous work on the SMC and LMC shows that the \hi is steeper than the dust power spectra we find here.
In the SMC, the \hi power spectrum is well-described by a single power-law (over the entire galaxy) with an index of $3.04\pm0.02$,
and between $5/3$ and $8/3$ \citep{Elmegreen2001ApJ...548..749E}, respectively.
\citet{Combes2012A&A...539A..67C} similarly find a steeper \hi power spectrum ($2.4\pm0.2$) relative to the IR bands and CO in M33.

To further this comparison, we calculate the \hi and CO integrated intensity power spectra of M33 and M31\footnote{We assume optically-thin \hi emission, but see \citet{Braun2012ApJ...749...87B}.}.
We use the Karl G. Jansky Very Large Array (VLA) M33 \hi map from \citet{Koch2018MNRAS} and IRAM 30-m \cotwoone map from \citet{Druard2014A&A...567A.118D}, both of which are convolved to the $167$~pc ($46.3\arcsec$) resolution of the dust surface density map.
For M31, we use a new VLA \hi map (Koch et al. in prep) with $58\arcsec$ resolution and the \coonezero map from \citet{Nieten2006A&A...453..459N}.
The M31 \hi map has a lower resolution than the dust and \co maps, so we convolve these maps to match the \hi. 
This gives a resolution of 201~pc ($58\arcsec$).

\begin{figure}
\includegraphics[width=0.5\textwidth]{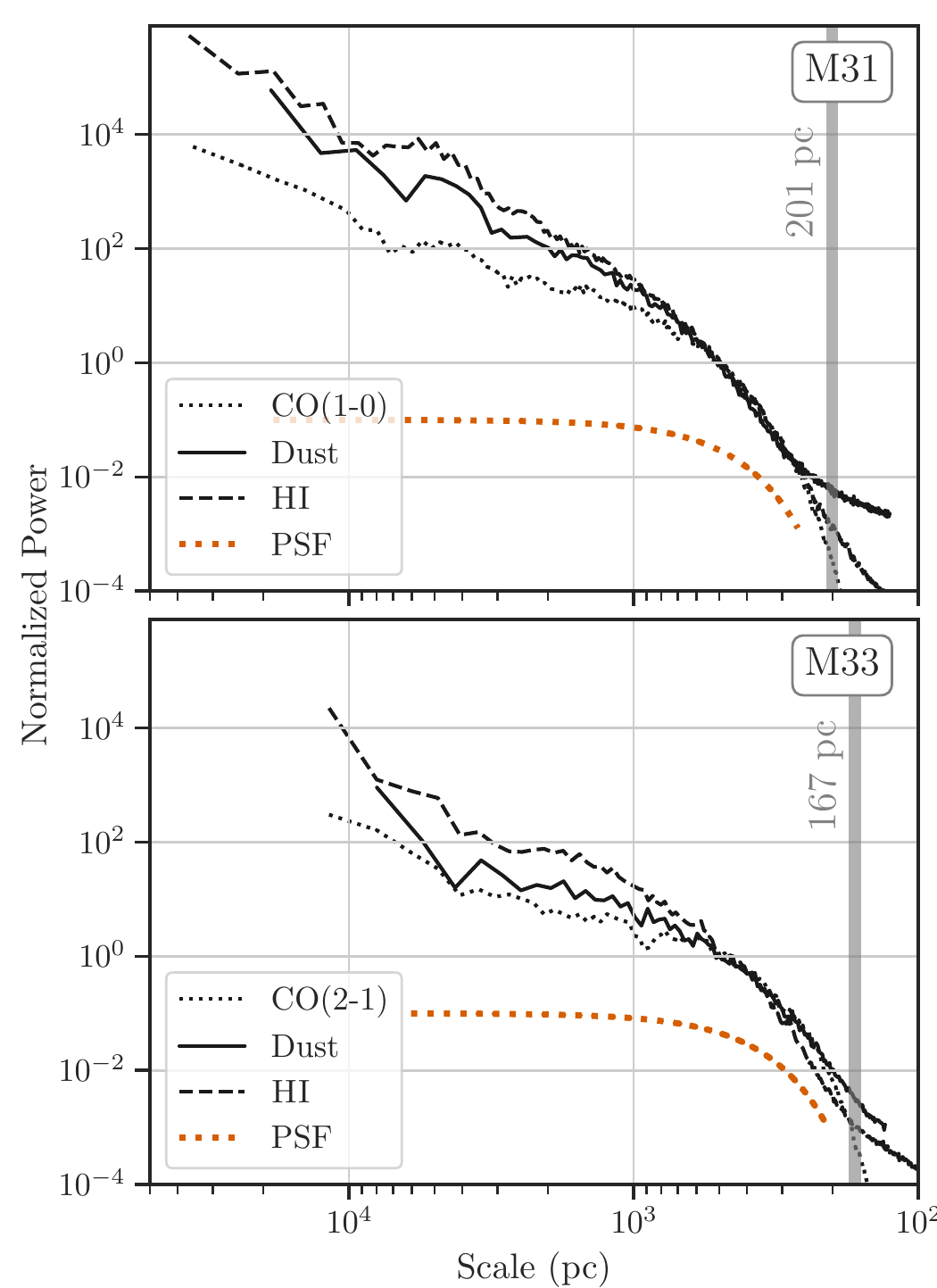}
\caption{\label{fig:m33_hi_dust_pspec} power spectra from CO, dust and \hi in M31 ($201$~pc;$58\arcsec$) and M33 ($167$~pc; $41\arcsec$) convolved to a common beam size and normalized to the power on $500$~pc scales. The power spectrum of the beam is shown with the orange dotted curve. All of the power spectra are well fit by a single power law attenuated by the beam. Indices are given in Table \ref{tab:hi_co_dust_indices}.  In both galaxies, the CO power spectrum is the flattest and the \hi is the steepest, with the dust somewhere in between.  In M33, the centralized \htwo emission significantly affects the dust power spectrum shape by comparison to the CO power spectrum.  This comparison shows that the power spectrum shape is affected by the large-scale structure of the emission on all measurable scales, making connections to 2D turbulence tenuous.}

\end{figure}

We fit the \hi and \co power spectra to Equation \ref{eq:obs_model}. Table \ref{tab:hi_co_dust_indices} presents the fit parameters for the \hi, \co, and dust power spectra.  The power spectra are all well fit by a single power-law model (Eq. \ref{eq:model}).  Figure \ref{fig:m33_hi_dust_pspec} shows the consistent differences between the \hi, dust, and \co power spectra, where \hi is the steepest and \co is the shallowest.
Our fits to the M33 \hi and \co are flatter than those found by \citet{Combes2012A&A...539A..67C}, though they use a resolution of $48$~pc for both and do not account for the PSF shape; fitting the \hi power spectrum at its native $80$~pc resolution gives an index of $2.30\pm0.13$, consistent with \citet{Combes2012A&A...539A..67C}.  \citet{Combes2012A&A...539A..67C} also include the smallest frequency bins from the power spectrum, which strongly deviate above the power-law relation. We exclude these data in our analysis.

A steeper \hi power spectrum relative to the dust implies a lack of power on small scales in the \hi.  In terms of the molecular and atomic column density, there are four sources for this discrepancy: (1) the \htwo distribution differs from the \hi on galactic scales, (2) saturated \hi on small scales, (3) optically-thick \hi dominates on small scales, and (4) radial changes in the dust-to-gas ratio due to radial metallicity gradient. The first three will remove structure on small-scales, smoothing the spatial distribution of \hi relative to the total gas distribution traced by dust.  The first two sources arise from the conversion of \hi to \htwo \citep[e.g.,][]{Bigiel2008AJ....136.2846B,Krumholz2013MNRAS.436.2747K,Sternberg2014ApJ...790...10S}. In M33, the \htwo distribution is centrally-concentrated in the galaxy and primarily from GMC scale emission \citep{Roso2003ApJ...599..258R}.  \citet{Koch2019MNRAS.485.2324K} show that the distribution of GMCs in M33 can provide an excess in power on scales up to $\sim2$~kpc, and can therefore affect the CO and dust power spectra shape on similar scales.  The final fourth point will tend to flatten the dust power spectrum on large scales, as the dust abundance decreases with metallicity.

The \htwo distribution in M31 is dominated by the ring-structures and does not show a strong concentration in the inner disc. However, with bright CO still clustered into a large-galactic structure (i.e., the rings), the morphology may still provide excess power on the scale of the rings, affecting the large-scale power spectrum.

We test the influence of the \htwo distribution on the dust surface density power spectrum by combining the \hi and CO maps to get the neutral gas surface density power spectrum. We assume constant Milky-Way $\alpha_{\rm CO}$ factor of $4.8$~\msolpcsq and $6.7$~\msolpcsq for the 1-0 and 2-1 lines, respectively, and optically-thin \hi to convert the integrated intensities to the molecular and atomic surface densities. All of the neutral gas surface density power spectra are well fit by a single power-law (Eq. \ref{eq:model}) with indices between the \hi and CO (Table \ref{tab:hi_co_dust_indices}).  We expect this result from the relative differences in the power spectra shown in Figure \ref{fig:m33_hi_dust_pspec}.

The power-law indices from the neutral gas power spectra are similar to the dust index, though the index is highly sensitive to the choice of $\alpha_{\rm CO}$, as shown in Table \ref{tab:hi_co_dust_indices}. For example, doubling $\alpha_{\rm CO}$ changes the index by $0.23$ and $0.30$ in M31 and M33, respectively.
We note that this is an overly-simplified treatment of $\alpha_{\rm CO}$ that only demonstrates the power spectrum's sensitivity to these variations.
The variations in the dust-to-gas ratio, tied primarily to metallicity \citep[e.g.,][]{Bolatto2013ARA&A..51..207B}, will show a similar sensitivity to changes in $\alpha_{\rm CO}$.
Both quantities also vary within galaxies \citep[e.g.,][]{SandstromLeroy2013ApJ...777....5S}, and a more sophisticated handling for these variations may explain the moderate discrepancies in the dust and gas power spectra.
Though the turbulent properties may change in the atomic and molecular ISM \citep{RomeoBurkert2010MNRAS.407.1223R} and dust may have different dynamics than the gas \citep{Hopkins2014ApJ...797...59H}, these variations in the netural gas power spectrum index suggest that conversion factors alone can explain the difference between dust and the gas phases.

\begin{table*}
    \caption{\label{tab:hi_co_dust_indices} Power spectrum indices of \hi, CO, dust, and total neutral gas surface density in M31 and M33, with different assumed values of $\alpha_{\rm CO}$.  In all cases, the power spectra are well-fit by a single power-law (Figure \ref{fig:m33_hi_dust_pspec}).  The M31 maps are convolved to the 197~pc ($57.5\arcsec$) beam from the \hi data (Koch et al. in prep.) and the M33 maps are convolved to the 167~pc ($41\arcsec$) beam of the dust surface density map.  Despite differences in the M31 and M33 indices, there is a consistent trend for steeper atomic gas power spectra and flatter molecular gas power spectra, with the dust power spectra somewhere in the middle.  The neutral gas surface density maps have power spectra similar to the dust, but are sensitive to the choice of $\alpha_{\rm CO}$. Differences between power spectra of different tracers emphasises how the large-scale structure affects the power spectrum shape on all measurable scales.  $^{\rm a}$ \coonezero \citep{Nieten2006A&A...453..459N}. $^{\rm b}$ \cotwoone \citep{Gardan2007A&A...473...91G,Gratier2010A&A...522A...3G,Druard2014A&A...567A.118D}.}
    \centering
    \begin{tabular}{ccc}
 & M31 & M33  \\  
 & 201 pc scales & 167 pc scales \\ \hline
$I_{\rm HI}$        &  $2.66\pm0.12$  & $2.18\pm0.10$ \\  
$I_{\rm CO}$        &  $1.59\pm0.08^{\rm a}$  & $0.91\pm0.14^{\rm b}$ \\
$\Sigma_{\rm dust}$ &  $2.44\pm0.15$  & $1.11\pm0.14$ \\
$\Sigma_{\rm HI} + \alpha_{\rm CO} I_{\rm CO}$ & $2.34\pm0.11$  & $1.66\pm0.12$ \\ 
    & $\alpha_{\rm CO10}=4.8$~M$_{\odot}\,{\rm pc^{-2}/\,\rm K\,km\,s^{-1}}$ & $\alpha_{\rm CO21}=6.7$~M$_{\odot}\,{\rm pc^{-2}/\,\rm K\,km\,s^{-1}}$ \\
$\Sigma_{\rm HI} + \alpha_{\rm CO} I_{\rm CO}$ & $2.11\pm0.10$  & $1.36\pm0.07$ \\
    & $\alpha_{\rm CO10}=9.6$~M$_{\odot}\,{\rm pc^{-2}/\,\rm K\,km\,s^{-1}}$ & $\alpha_{\rm CO21}=13.4$~M$_{\odot}\,{\rm pc^{-2}/\,\rm K\,km\,s^{-1}}$ \\
\end{tabular}
\end{table*}

In all cases, we do not correct for optically-thick \hi when computing the \hi surface density as, for M33, \citet{Koch2018MNRAS} do not find evidence for flattened \hi velocity spectra indicative of bright optically-thick \hi emission \citep{Braun1997ApJ...484..637B,Braun2009ApJ...695..937B}, though \citet{Braun2012ApJ...749...87B} argue there is a 30\% correction factor to the atomic gas mass in M31 and M33.  If optically-thick \hi emission contributes to the lack of power in the \hi power spectrum at small scales, we expect it to arise from spatial regions $<167$~pc in size.  We note, however, that \citet{NestingenPalm2017ApJ...845...53N} find no change in the SMC \hi power spectrum when correcting for optically-thick \hi \citep{Stanimirovic1999MNRAS.302..417S}, and \citet{Pingel2018ApJ...856..136P} also find no change in the index for \hi of the Perseus molecular cloud.

These results demonstrate that the power spectrum of the dust, and IR bands dominated by dust emission, is strongly influenced by the location of \htwo, in this case traced by CO, leading to a significantly different slope relative to only the atomic component traced by the \hi. These differences in the power spectra of the atomic and molecular power spectra, and between different galaxies, strongly suggests that the properties of the large-scale power spectrum are dominated by the galactic distribution of the tracer.  This makes comparisons to 2D turbulent properties on $>$~kpc scales tenuous without accounting for these differences.

\subsection{Power spectrum breaks are not ubiquitous}
\label{sub:disc_scale_height}

A key result from previous studies using the spatial power spectrum of dust and gas in nearby galaxies is a break in the power spectrum on scales similar to the expected disc scale height, which is otherwise difficult to constrain from observations of low or moderately inclined galaxies. Here we find that previous claims of a power spectrum break for M33 \citep{Combes2012A&A...539A..67C} and the LMC \citep{Block2010ApJ...718L...1B} can be entirely accounted for by the shape of the PSF. The only exception we find is the 24~$\mu$m LMC map, where 30 Doradus is sufficiently bright to cause an excess on $\sim200$~pc scales (\S\ref{fig:lmc_mips24_30dor}). In \S\ref{sub:does_the_power_spectrum_index_change_on_}, we find that splitting the map does not make a power spectrum break evident, which may occur if the disc scale height changes substantially over the maps, thereby smearing out a single clear break-point.  With the lack of a power spectrum break, the spatial power spectrum does not constrain the disc scale height.

Most studies that explore the power spectrum break find that it is located on scales a few times the PSF FWHM \citep{Elmegreen2001ApJ...548..749E,Dutta2009MNRAS.397L..60D,Block2010ApJ...718L...1B,Combes2012A&A...539A..67C}. Based on our results, this suggests that the scale of the break could be influenced by the PSF shape of the observation. A similar suggestion is made by \citet{Grisdale2017MNRAS.466.1093G} based on \hi power spectra of 6 galaxies from THINGS \citep{Walter2008AJ....136.2563W}.  There are some exceptions where breaks are found on scales many times the PSF FWHM, however, these tend to be measured at either $24$~$\mu$m \citep{Block2010ApJ...718L...1B} or the H$\alpha$ line \citep{Willett2005AJ....129.2186W,Combes2012A&A...539A..67C} where a small number of giant \hii regions provide a significant fraction of the total flux of the galaxy.

%

\citet{SzotkowskiYoder2019arXiv191104370S} have recently found power spectrum breaks in the \hi in the LMC when measured over local scales.  Several areas in their analysis show a break on scales significantly larger than the beam size ($30$~pc), with variations around near regions with strong stellar feedback (e.g., giant \hii regions). This is in apparent disagreement with the lack of break points we find in the local LMC dust surface density power spectra (\S\ref{sub:does_the_power_spectrum_index_change_on_}), yet we show in \S\ref{sub:comparisons_with_other_ism_tracers} that dust and \hi power spectra are different when measured over the entirety of M31 and M33. These differences could indicate that the \hi, which saturates above some surface density \citep[e.g.,][]{Krumholz2013MNRAS.436.2747K}, better traces the influence of stellar feedback on the surrounding atomic ISM.  There is significant precedent for feedback affecting galaxy-scale spatial power spectra from numerical studies \citep{Bournaud2010MNRAS.409.1088B,Pilkington2011MNRAS.417.2891P,Combes2012A&A...539A..67C,Grisdale2017MNRAS.466.1093G}, including those that do not find a power spectrum break \citep{Renaud2013MNRAS.436.1836R}.

These results point to multiple factors that influence the power spectrum shape and the presence of a break, rather than a ubiquitous break related to the disc scale height. These factors include the large-scale distribution of gas in the galaxy, especially the presence of high column density, \htwo-dominated regions (\S\ref{sub:galaxy_variations}), and the gas tracer used (\S\ref{sub:comparisons_with_other_ism_tracers}).  The relative influence of each factor can be explored using local power spectra (\S\ref{sub:does_the_power_spectrum_index_change_on_}) of multiple tracers.  We plan to explore this in future work.

We also note that power spectrum breaks are not commonly found on smaller scales within nearby Milky Way molecular clouds ($<20$~pc).  Power spectrum studies of the Perseus molecular cloud include scales where stellar feedback provides sufficient energy to drive turbulence \citep{Padoan2009ApJ...707L.153P,Arce2011ApJ...742..105A} but do not find a power spectrum break, despite results from alternative methods, like the probability distribution function (PDF), that suggest small scale driving should be dominant \citep{Bialy2017ApJ...843...92B}.

The multiple factors influencing the power spectrum shape do not rule out a break at the disc scale height, tracing the transition from 3D to 2D turbulence.  It may be possible to account for each of these factors with a more sophisticated model to search the uniform presence of a break.  However, the current quality of data does not support the need for a more complex model.

Finally, we note that this analysis is limited to information from the projected density field of these galaxies.  When using a spectral-line, the line-of-sight velocity offers additional information useful for this type of analysis.  Velocity information be incorporated into the power spectrum or structure function by using the line-of-sight velocity centroid \citep[e.g.,][]{Bertram2015MNRAS.446.3777B} or different on power spectra of the whole spectral-line data cube, such as the Velocity Channel Analysis or Velocity Coordinate Spectrum \citep[e.g.,][]{Stanimirovic2001ApJ...551L..53S,LazarianPogosyan2006ApJ...652.1348L,ChepurnovBurkhart2015ApJ...810...33C}.
Alternatively, empirically-based methods like the Spectral Correlation Function \citep[SCF;][]{Rosolowsky1999ApJ...524..887R} provide a complementary measure of structure with spatial scale. \citet{Padoan2001ApJ...555L..33P} found deviations in the SCF relation applied to \hi data of the LMC and attributed the deviations to the LMC disc scale height.
In future work, we will utilize velocity information in our analysis of \hi of M33 \citep{Koch2018MNRAS} and M31 (Koch et al.~in prep.).

\section{Summary}
\label{sec:summary}

We present a unified analysis of the 1D power spectra of mid- to far-IR emission and the dust surface density in the LMC, SMC, M31, and M33. A key result of our work is that previous claims of a power spectrum break can be explained by the instrumental PSF response and are not a measurement of the disc scale height.  This result has important consequences for simulations of Local Group-like galaxies, which have also found break points in power spectra \citep[e.g.,][]{Bournaud2010MNRAS.409.1088B,Combes2012A&A...539A..67C,Grisdale2017MNRAS.466.1093G}.

\begin{enumerate}
    \item We model the PSF response on the 1D spatial power spectra and find that the power spectra of all the galaxies is well-modeled by a single power-law plus point source components. We demonstrate that previous studies that find a break point in the power spectra is entirely due to the PSF response.  We also note that, comparing to both Galactic and extra-galactic power spectrum studies, there are few cases a power spectrum break from intensity maps is unambiguously found over all spatial scales across several wavebands.
    
    \item M31, the LMC, and SMC have similar power spectra indices ranging from $2$ to $2.5$. The indices in these three galaxies are broadly consistent in the dust surface density and individual infrared bands despite the difference in their morphology.  This similarity demonstrates that different spatial morphologies can produce similar power spectra, showing the need to carefully consider multiple sources that can alter the power spectrum shape. 
    
    \item Compared to the other three galaxies, M33 has a significantly flatter power spectrum with an index of $\sim1.3$.  We calculate the power spectrum of IC 342, a nearby face-on spiral, in the {\it Herschel} bands and find a similarly flat power-law index.  This similarity suggests that spiral galaxies with a central \htwo concentration tend to have flatter power spectra, which can be explained by the shape of an exponential (molecular) disc with a flat (mostly atomic) component, rather than large-scale 2D turbulence.
    
    \item We compare the dust, \hi, and CO power spectra of M31 and M33 at a common scale. The \hi and CO power spectra are well-fit by a single power-law. We find a consistent trend in the indices, with \hi being the steepest and CO being the shallowest.  This is consistent with \hi having more structure on large scales and CO having more structure on small scales.  We create total neutral gas surface density maps by combining the \hi and CO,
    and find their power spectrum index is intermediate between the \hi and CO, and is similar to the dust.
    The neutral gas power spectrum is sensitive to $\alpha_{\rm CO}$, leading to variations that can account for the difference in the power spectra index of the dust.
    The dust and gas are further related by the dust-to-gas ratio, which is known to vary on large scales \citep{SandstromLeroy2013ApJ...777....5S}.
    This result provides further evidence that the power spectrum is sensitive to the large-scale distribution of a tracer, making it difficult to connect to 2D turbulence without accounting for this effect.
    
    \item We compute the dust surface density power spectra over local ($\sim820$~pc) regions within the LMC and SMC and find they are also well-fit without a power spectrum break. This result rules out local variations in the disc scale height as an explanation for the lack of a break measured from the whole galaxy's power spectrum. The difference between the dust and \hi power spectra that we find, and the recent identification of local \hi breaks in the LMC by \citet{SzotkowskiYoder2019arXiv191104370S} shows that the \hi may better trace feedback relative to the total neutral gas column traced by the dust.
    
    \item The local power spectra in the LMC and SMC show substantial variation across the galaxies.  We find that steeper power spectra occur near large intensity gradients, similar to what \citet{Burkhart2010ApJ...708.1204B} find using the skewness and kurtosis of the \hi column density in the SMC.

    \item Simulations of Local Group-like galaxies reflect some of the results we find here. \citet{Grisdale2017MNRAS.466.1093G} find that the column density power spectrum on $\sim$few kpc scales is sensitive to the mass distribution in galaxies but is insensitive to other effects like stellar feedback. They also find that, when comparing to \hi intensity power spectra of nearby galaxies, the small-scale power spectra are dominated by the PSF shape. These results demonstrate the need to produce synthetic observations when comparing power spectra of simulations and observations \citep{Haworth2018NewAR..82....1H}. Our results provide a benchmark for comparing observations and simulations of Local Group-like galaxies.
    
\end{enumerate}

Our results demonstrate that power spectra are sensitive to systematic effects that significantly effect how they are modeled and interpreted. Where applicable, we recommend forward-modelling the instrument PSF when fitting a model to power spectra.

Previous work has focused on the source of the power spectrum break, using it as a measure of the disc scale height to constrain galactic structure. With these results, we show that an alternative explanation is required to understand the ubiquity of galactic power spectra on scales well below the disc scale height. Further work requires investigating whether a break is measurable from the velocity field from spectral lines \citep[e.g.,][]{Padoan2001ApJ...555L..33P}.

Scripts to reproduce the analysis are available at \url{github.com/e-koch/DustyPowerSpectra}\footnote{Code DOI: \url{https://doi.org/10.5281/zenodo.3583220}}.

\section*{Acknowledgments}

We thank the referee, Alessandro Romeo, for helpful comments that improved manuscript.
EWK acknowledges helpful discussions with Bruce Elmegreen and Sne\v{z}ana Stanimirovi\'{c}.
EWK is supported by a Canada Graduate Scholarship and Michael Smith Foreign Study Supplement from the Natural Sciences and Engineering Research Council of Canada.  EWK and EWR are supported by a Discovery Grant from Natural Sciences and Engineering Research Council of Canada (RGPIN-2017-03987).
The work of DU and AKL is partially supported by the National Science Foundation under Grants No. 1615105, 1615109, and 1653300.
{\it Herschel} is an ESA space observatory with science instruments provided by European-led Principal Investigator consortia and with important participation from NASA. This work is based, in part, on observations made with the {\it Spitzer Space Telescope}, which is operated by the Jet Propulsion Laboratory, California Institute of Technology under a contract with NASA.
The National Radio Astronomy Observatory and the Green Bank Observatory are facilities of the National Science Foundation operated under cooperative agreement by Associated Universities, Inc.

\textbf{Code Bibliography: }
astropy \citep{astropy,astropy2} --- radio-astro-tools (spectral-cube, radio-beam; \url{radio-astro-tools.github.io}) --- matplotlib \citep{mpl} ---  seaborn \citep{seaborn} --- corner \citep{corner} --- pymc3 \citep{pymc3} --- numpy \& scipy \citep{numpy_oliphant2006guide} \\

\bibliographystyle{mn2e}
\bibliography{ref}

\appendix

\section{Additional systematics affecting fits}
\label{app:additional_systematics}

Some of the fits to the power spectra presented in Table \ref{tab:band_fits} required an altered model or additional step applied to the data to find a valid fit. These cases are shown in Table \ref{tab:coldens_fits} with an additional symbol. We provide further details of these special cases here. 

\begin{itemize}
    \item SMC MIPS 70~$\mu$m -- There is cross-hatching in the map on scales below the PSF's FWHM. Thus, the power spectrum does not follow the PSF on small scales. We do not use the PSF in the model and restrict the fitting to scales above $25$~pc.
    \item SMC MIPS 160~$\mu$m -- There are large, noisy values at the edge of these images that cause ringing in the FFT. We apply a Tukey apodizing kernel to taper these values in the power spectrum.
    \item SMC PACS 160~$\mu$m -- We do not use the PSF for the fit for the same reason as the MIPS 70~$\mu$m. The fitting is restricted to scales larger than $10$~pc.  We also apply a Tukey apodizing kernel to this map, similar to the MIPS 160~$\mu$m map.
    \item M31 PACS 100~$\mu$m -- We do not use the PSF in the fit as the small-scales in the power spectrum are dominated by the scan pattern. We restrict the fitting to scales larger than $150$~pc.
    \item M31 MIPS 160~$\mu$m, PACS 160~$\mu$m, SPIRE 250, 350, 500~$\mu$m -- We apply a Tukey apodizing kernel to taper large values at the map edge, avoiding ringing in the FFT.
\end{itemize}

We also test whether bright foreground point sources affect the power spectrum shape or contribute to the additional systematics described above. Using the MIPS point-source subtracted maps from the SAGE \citep{Meixner2006AJ....132.2268M} and SAGE-SMC \citep{Gordon2006ApJ...638L..87G,Gordon2011AJ....142..102G} data releases, we find no difference in the power spectrum index. We do not expect point source contamination to affect our results.

\section{Additional Figures}
\label{app:additional_figures}

We provide figures of the power spectra for each band, as shown in Figure \ref{fig:coldens_fits}, as part of the supplemental paper figures. The fitted values are given in Table \ref{tab:band_fits}.




\section{Deprojection does not change the large-scale power spectrum}
\label{app:deprojection}

Some previous works presenting extragalactic spatial power spectra have deprojected the image into the galaxy frame prior to computing the power spectrum \citep{Block2010ApJ...718L...1B,Combes2012A&A...539A..67C}. We show an example of deprojection using the SPIRE 500~$\mu$m maps of M31 and M33 to demonstrate that deprojection does not significantly affect the large-scales of the power spectrum.  We use these maps for this example because the SPIRE 500~$\mu$m map have well-behaved PSFs with near-uniform noise, which allows for the deprojected PSF to be well-described by an elliptical Gaussian.

We use ths position angle and inclination for M31 ($i=77.7\degree$;${\rm PA}=38\degree$) and M33 ($i=55.1\degree$;${\rm PA}=201\degree$) from \citet{Corbelli2010A&A...511A..89C} and \citet{Koch2018MNRAS}, respectively. We deproject each map in three steps: (1) the galaxy centre is shifted to the central image pixel, (2) the image is rotated to have the semi-major axis aligned along the y-axis of the image, and (3) the image is warped along the minor axis to match the major axis. Each of these steps are applied with interpolation methods in the {\sc scipy.ndimage} package\footnote{\url{docs.scipy.org/doc/scipy/reference/ndimage.html}}. For a Gaussian PSF, step (3) can be applied to the PSF to approximate the PSF in the deprojected frame; thus the deprojected images have a larger effective beam shape set by the inclination.

\begin{figure*}
\includegraphics[width=0.9\textwidth]{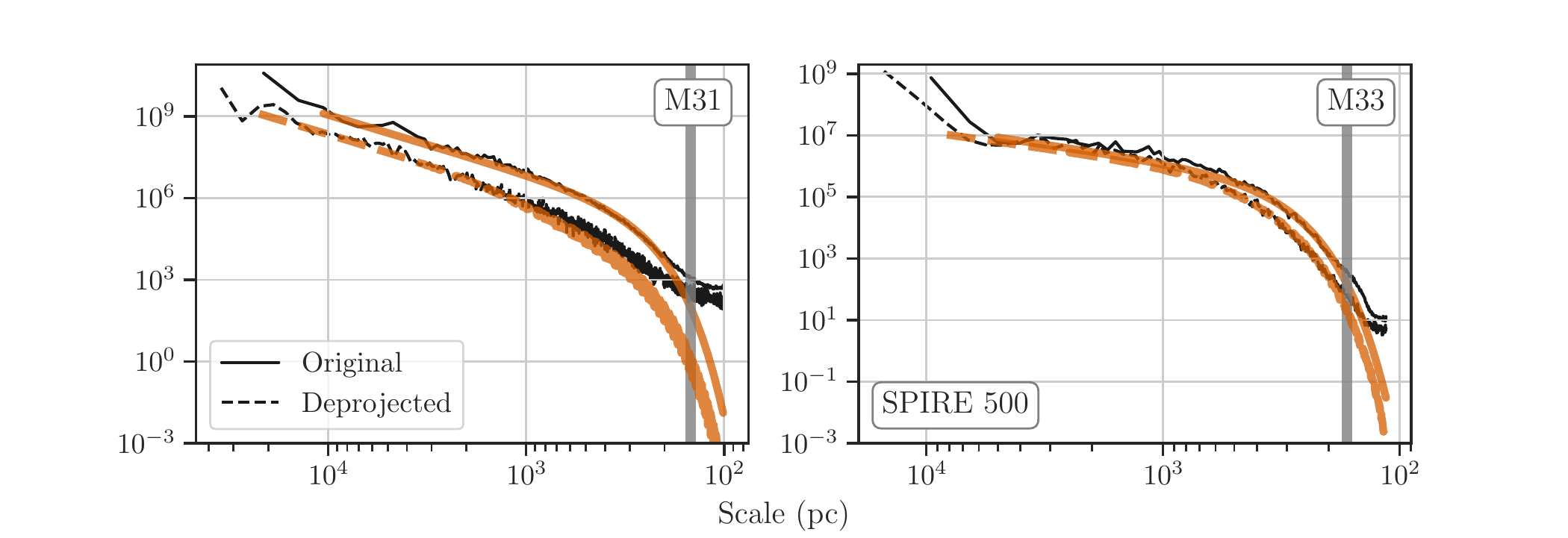}
\caption{\label{fig:deproj_example} Comparison of the M31 and M33 SPIRE 500~$\mu$m power spectra with and without deprojecting the maps. The translucent orange lines show the best fit to Equation \ref{eq:obs_model} and the vertical lines are the FWHM of the original image.  The warping step for deprojection effectively adds a ``white noise'' due to the interpolation that is evident on scales below the FWHM. This is more evident for M31 due to its large inclination. We limit the fit to scales larger than $3\times{\rm FWHM}$ to avoid these regions.  For M31, the deprojected power spectrum index is $2.20\pm0.19$, consistent with the original power spectrum index of $2.17\pm0.12$. Similarly, the M33 deprojected power spectrum index $1.08\pm0.08$ is consistent with the original index of $1.15\pm0.08$. Thus, we expect that deprojection will not significantly alter the fitted indices in this paper.}
\end{figure*}

Figure \ref{fig:deproj_example} shows the original and deprojected power spectra for M31 and M33.  In both cases, the shape of the power spectrum on large scales (small frequencies) is not affected. The fitted indices to the deprojected power spectra are $2.17\pm0.12$ and $1.08\pm0.08$ for M31 and M33, respectively; these indices are consistent with the original power spectra indices of $2.17\pm0.12$ and $1.15\pm0.08$ (Table \ref{tab:band_fits}).

The power spectrum index after deprojection is consistent with \citet{Grisdale2017MNRAS.466.1093G}, who compare power spectra of simulated galaxy discs at $i=0\degree$ and $i=40\degree$ and find that only scales of order the disc diameter are affected by inclination. \citet{Block2010ApJ...718L...1B} also note no significant difference in indices from deprojecting the MIPS maps of the LMC.

Finally, we note the difficulty in simultaneously modelling for instrumentation effects and the projection effects from the observed frame of the galaxy. Deprojection of a non-axisymmetric PSF is complicated by the rotation step. Fully modelling for both of these effects would require forward-modelling the 2D power spectrum through a deprojection step, followed by applying the PSF. The computational requirements to model the 2D power spectrum for large images would be prohibitive in practice.

\section{Local LMC and SMC power spectrum uncertainty}
\label{app:broken_plaw_model}

\begin{figure*}
\includegraphics[width=\textwidth]{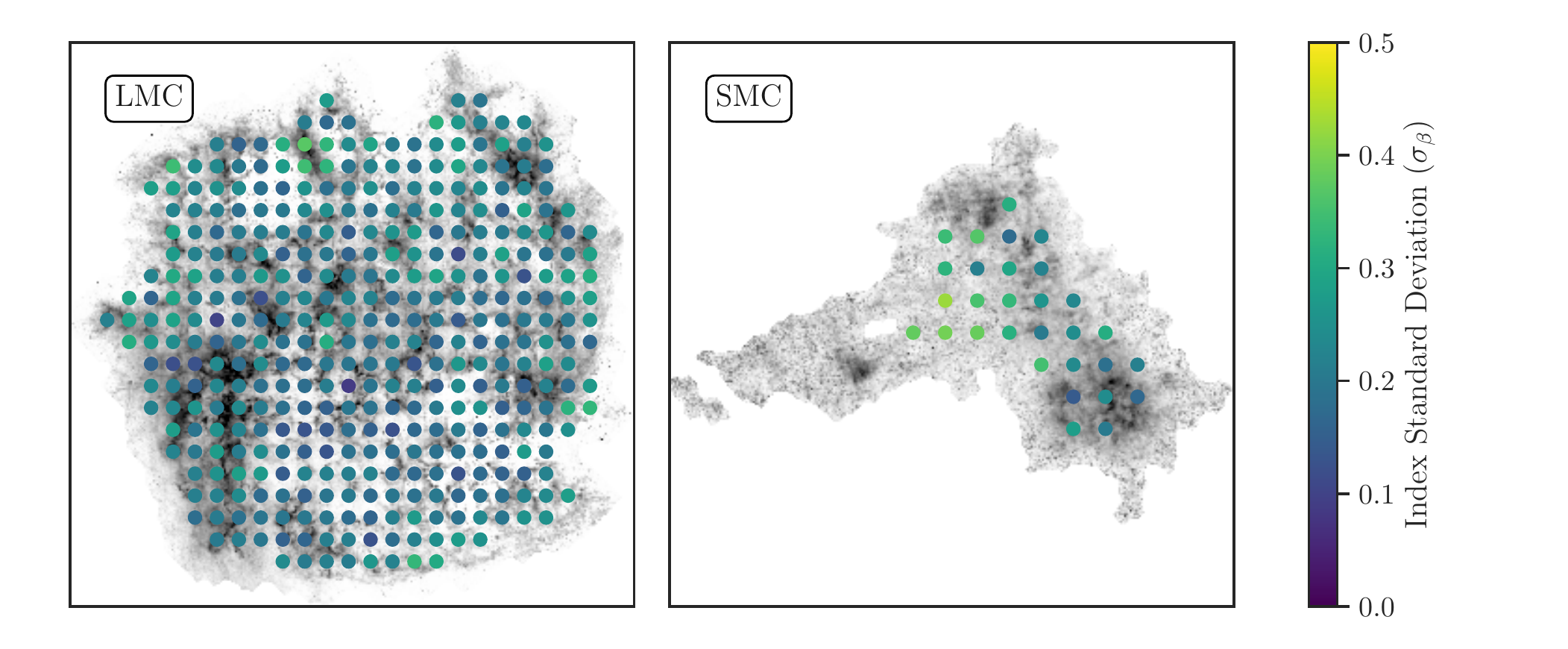}
\caption{\label{fig:magcloud_index_spatvar_errs} The power spectrum index uncertainties in $820$~pc regions overlaid on the LMC and SMC dust surface density.  The power spectrum indices are shown in Figure \ref{fig:magcloud_index_spatvar}.}
\end{figure*}

Figure \ref{fig:magcloud_index_spatvar_errs} shows the power spectrum index uncertainties for the values shown in Figure \ref{fig:magcloud_index_spatvar}.  The uncertainties are small relative to the change in the indices, indicating that spatial variation in Figure \ref{fig:magcloud_index_spatvar} represent real variations in the power spectrum shape.

\section{Dust power spectra of IC342}
\label{app:ic342}

In \S\ref{sub:galaxy_variations}, we perform a similar analysis to \S\ref{sec:power_spectra} on the {\it Herschel} maps of IC342 to compare with M33's power spectra. Despite being $4\times$ the distance of M33, IC342 is one of the nearest face-on spiral galaxies. Critically for this comparison, the molecular gas fraction increases towards the inner disc, similar to M33 and other spiral galaxies but unlike the LMC, SMC, and M31.

Similar to the Local Group galaxies, we calculate the power spectrum centered on IC342 and exclude regions far from the galaxy.  This step is more critical for IC 342 than the other galaxies because of its low Galactic latitude; Galactic emission in the {\it Herschel} bands is substantial over most of the maps.  We do not find substantial contamination from Galactic emission over the regions used for the power spectrum, which would be indicated by emission near the edges of the region causing the Gibbs phenomenon in the 2D power spectrum.

Table \ref{tab:ic342_fits} provides fit results to Equation \ref{eq:obs_model} for the {\it Herschel} bands, excluding the PACS 100~$\mu$m map due to variations from the expected PSF shape that appear to be systematics (Appendix \ref{app:additional_systematics}). In \S\ref{sub:galaxy_variations}, we compare these results with the other galaxies.


\begin{table*}
    \caption{\label{tab:ic342_fits} Fit parameters to Equation \ref{eq:obs_model} for the {\it Herschel} bands of IC 342. Uncertainties are the 1-$\sigma$ interval estimated from the MCMC samples. None of the fits constrain the unresolved point-source term $B$. We exclude the PACS 100~$\mu$m map as the small scale structure does not follow the expected shape of the PSF (see Appendix \ref{app:additional_systematics}). IC 342 has a flat power spectrum similar to M33, suggesting that a flatter power spectrum is associated with typical spirals where the molecular gas fraction increases in the inner disc, unlike the LMC, SMC, and M31.}
    \centering
    \begin{tabular}{ccccc}
Band & Resolution ($\arcsec$) & Phys. Resolution (pc) & log$_{10}$ $A$ & $\beta$ \\ \hline 
PACS 160  &  11.2 &  181 & $2.49\pm0.04$ & $0.99\pm0.05$  \\  
SPIRE 250 &  18.2 &  295 & $8.03\pm0.07$ & $1.04\pm0.06$  \\  
SPIRE 350 &  25   &  405 & $6.61\pm0.10$ & $0.98\pm0.10$  \\  
SPIRE 500 &  36.4 &  589 & $5.30\pm0.16$ & $0.89\pm0.17$  \\  
 
\end{tabular}

\end{table*}

\label{lastpage}
\end{document}